\documentclass{article}

\usepackage{amssymb}

\def\draft{draft}
\def\final{final}

\def\mode{final}

\usepackage{graphicx}
\usepackage{epsfig}
\newcommand{\beq}{\begin{equation}}
\newcommand{\eeq}{\end{equation}}
\newcommand{\beqa}{\begin{eqnarray}}
\newcommand{\eeqa}{\end{eqnarray}}
\newcommand{\beqar}{\begin{eqnarray*}}
\newcommand{\eeqar}{\end{eqnarray*}}

\newcommand{\labell}[1]{\label{#1}\qquad_{#1}} %{\label{#1}}
\newcommand{\reef}[1]{(\ref{#1})}

\newcommand{\eg}{{\it e.g.,}\ }
\newcommand{\ie}{{\it i.e.,}\ }

\newcommand{\norm}[1]{\raise.3ex\hbox{:}#1\raise.3ex\hbox{:}}

\renewcommand\d{{\rm d}}
\newcommand\prt{\partial}

\newcommand{\gsim}{\mathrel{\raisebox{-.6ex}{$\stackrel{\textstyle>}{\sim}$}}}

\renewcommand{\b}{\beta}

\newcommand\cL{{\cal L}}

\renewcommand\Re{R_{\rm e}}      %<<warning!

\newcommand{\unit}[1]{\ \mathrm{#1}} 
\newcommand{\td}{\widetilde}
\newcommand{\half}{\frac{1}{2}}

\newcommand{\pdrvs}[2]{\frac{\partial^2 #1}{\partial #2^2}}

\newcommand{\ra}{\rightarrow}

\newcommand{\bi}{\begin{itemize}}
\newcommand{\ei}{\end{itemize}}
\newcommand{\ben}{\begin{enumerate}}
\newcommand{\een}{\end{enumerate}}

\newcommand{\const}{\mbox{const}}
\newcommand{\bmtx}{\left[ \begin{array}{cc}}
\newcommand{\emtx}{\end{array} \right]}
\newcommand{\bvec}{\left[ \begin{array}{c}}
\newcommand{\evec}{\end{array} \right]}

\newcommand{\mysection}[1]{\section{#1} \setcounter{equation}{0}}

\renewcommand{\Re}{{\mbox{Re }}}
\renewcommand{\Im}{{\mbox{Im }}}

\newcommand{\ZZ}{{\rm Z}}

\newcommand{\bfig}{\begin{figure}} 
\newcommand{\efig}{\end{figure}}
\newcommand{\nn}{\nonumber}
\newcommand{\ba}{\begin{array}}
\newcommand{\ea}{\end{array}}

\newcommand{\bs}{\begin{slide}}
\newcommand{\es}{\end{slide}}

\newcommand{\tdz}{\tilde z}

\renewcommand{\newline}{\\}

\ifx\mode\final \renewcommand{\labell}[1]{\label{#1}}\fi

\ifx\mode\draft \newcommand{\mybibitem}[1]{\bibitem[#1]{#1}}
\else\newcommand{\mybibitem}[1]{\bibitem{#1}}\fi

\title{On Tunnelling In Two-Throat Warped Reheating}
\author{Peter Langfelder \newline
\textit{Dept. Of Physics, University of Waterloo and Perimeter
Institute} \newline
\textit{Waterloo, ON, Canada } \newline
\texttt{plangfel@uwaterloo.ca} 
}
\date{\today }

\begin{document}

\setcounter{page}{0}

\maketitle

\begin{center}
\vspace{1cm}
{\large \bf Abstract} \\
\end{center}
\vspace{\baselineskip}
We revisit the energy transfer necessary for the warped reheating scenario in a two-throat geometry.
We study KK mode wavefunctions of the full two-throat system in the Randall--Sundrum (RS) approximation and find an
interesting subtlety in the calculation of the KK mode tunnelling rate.
While wavepacket tunnelling is suppressed unless the Standard Model throat is very long, 
wavefunctions of modes localized in different throats have a non-zero overlap and energy
can be transferred between the throats by interactions between such KK modes.
The corresponding decay rates are calculated and found to be faster than the tunnelling rates found in
previously published works. However, it turns out that the imaginary parts of the mode frequencies, induced by the decay, 
slow the decay rates themselves down. The self-consistent decay rate turns out to be given by the plane wave tunnelling
rate considered previously in the literature.
We then discuss mechanisms that may enhance the energy transfer between the throats over the
RS rates. In particular, we study models in which the warp factor changes in the UV region less
abruptly than in the RS model, and find that it is easy to build phenomenological models in which the plane wave
tunnelling rate, and hence the KK mode interaction rates, are enhanced compared to the standard
RS setup. 

%Typeset in mode \mode.

%\ifx\mode\draft This is a draft
   %\else This is final \fi

%\end{abstract}

\thispagestyle{empty}

\newpage

\tableofcontents

%\setcounter{section}{-1}
%
%\mysection{Still to do (to be emptied before publishing)}
%
%\bi
%
%\item emphasize differences: energy transfer without tunnelling, KK->2g decay possible
%
%\item Could do the same things with higher bessel functions representing KK modes with internal angular
%momentum
%
%\ei

\mysection{Introduction}

String theory provides several promising and in many ways natural avenues to address the origin and
properties of inflation as well as the origin of hierarchies of energy scales. 
For example, when the 6-dimensional internal space consists of several
(perhaps many) long throats joined by bulk~\cite{GKP}, physical processes deep in a particular throat
will have their energies redshifted by an amount that depends on their position in the
throat. In this manner one can obtain large hierarchies simply by cleverly choosing the
internal geometry and positions of various physically important objects within it~\cite{GKP}: for example, placing the
Standard Model on a stack of D branes deep inside a long throat can lead to a desired low SM energy
scale~\cite{RS1, verlinde9906}.

An interesting class of inflationary models in such a multi-throat background is driven by the potential energy of
D3 and anti-D3 branes~\cite{dvali9812}. While the D3 branes are mobile, the anti-D3 branes are stuck at the bottom of one of
the throats; during the ``slow-roll'' phase of inflation, the D3 branes move through the bulk and the
appropriate throat(s) toward the anti-D3 branes~\cite{chen0408, KKLMMT}. Once they get within about a string length, a tachyonic mode
appears and the branes and antibranes annihilate, ending inflation. The annihilation is expected to produce a
large number of very massive closed strings~\cite{llm} that
will quickly decay into the lowest string states~\cite{gary}, namely the (10-dimensional) graviton, the gauge fields and the
dilaton, and their fermionic partners. 

Typically the annihilating branes are
assumed to be located in a moderately short throat, so inflation happens at the scale required by CMB
observations, about
$10^{14}$--$10^{15} \unit{GeV}$~\cite{barnaby}. While the annihilation products have in
principle enough energy to reheat our universe, the energy must first be transferred to our vicinity in the
internal space, which usually means tunnelling through the bulk joining the throats. This issue received a
good deal of attention recently~\cite{gary, barnaby, kofman, rob, henry, henry0602}, with the general
conclusion being that generically the tunnelling seems to be too slow, but can be made fast enough for
certain specific choices of the parameters in the model. 

Our aim is to make more precise several qualitative arguments and rough estimates used
in the works cited above (the first attempt in this direction appeared in~\cite{gary}). Further, we investigate various
mechanisms that may enhance (or, in some cases, suppress) the rate of energy transfer between the
annihilation (A) and the the Standard Model (SM) throats. Following~\cite{gary, barnaby, kofman}, we will
(mostly) work with two 5-dimensional AdS throats of constant
(and equal) curvature separated by a UV brane (this setup can also be viewed as a two-fold copy of
the Randall--Sundrum model~\cite{RS1}, and we will occasionally refer to it as the doubled RS
model). 

We begin in Section~\ref{sec:background} by reviewing the setup of~\cite{barnaby,
dimopoulos0106} that will be the background for our tunnelling and decay calculations.
In Section~\ref{sec:Modes} we consider, in full detail, the 
wavefunctions of KK modes in this model and compare them to the wavefunction of the
4-dimensional graviton (zero mode). We find that the KK mode wavefunctions are enhanced only within narrow
``resonance'' bands; the width (in energy) of these bands can also be interpreted as the tunnelling rate from
the A throat into the SM one. In Section~\ref{sec:FiniteSMThroat} we find that the narrow width is
also responsible for the (perhaps surprising) fact that for tunnelling to actually happen, the SM throat must
be much longer than the A throat.

Even if KK modes localized in the A throat do not tunnel, their wavefunctions have a small tail in the SM throat. Thus
they can decay into KK modes localized in the SM throat. In Section~\ref{sec:Decay} we study the corresponding decay
rates.  Our results indicate that the decay
into KK modes is faster than the tunnelling rates calculated in~\cite{gary, barnaby, dimopoulos0106} and can dominate
even if the SM throat is not as long as would be required for tunnelling. However, it turns out that the fast decay
rates are not physical: in Section~\ref{sec:DecayEffects} we discuss modifications of the decay rate calculations due
to different decay rates in the A and SM throats.\footnote{We were motivated by~\cite{kofman} who argue that differing
decay rates can lead to a significant suppression of tunnelling.} We find that having different decay rates in the two
throats changes the shape of the wavefunction dramatically: the mode will penetrate the side on which it decays faster
only to a certain distance that is inversely proportional to the decay rate. This implies that long throats cannot
enhance decay rates indefinitely; in fact, it turns out that the self-consistent decay rate is limited by the plane wave
tunnelling rate considered in~\cite{gary, barnaby, dimopoulos0106}. We conclude that the plane wave tunnelling is
realized, but only because the tunnelled particles decay in the SM throat fast enough. 

Lest the reader thinks that this whole exercise simply re-derived a simple result in a complicated way, the
calculation of the self-consistent decay rate allows us to check that the tunnelling and decay are not slowed down
further by the effect of complex frequencies (this possibility was pointed out in~\cite{kofman}).

In Section~\ref{sec:LevelCrossing} we look at a possible mechanism for energy transfer that arises
during SM throat relaxation after inflation. Intuitively it can be understood as follows: as the length of
the SM throat increases, the spectrum of masses of its KK modes shifts down, while the A throat KK mode
masses remain unchanged. When an SM throat eigenvalue approaches an A throat eigenvalue, instead of
crossing they ``repel'' each other\footnote{This is known as level repulsion in standard Quantum Mechanics}:
what was a KK mode localized in the A throat becomes a mode localized in the SM throat and
vice-versa. Thus, KK modes and their energy could be ``sucked'' out of the A throat and into the SM
throat where they could decay before another level repulsion would push them back. Unfortunately it turns out
that the presence of an imaginary part of the mode mass (due to the decay) allows modes to avoid each other
in the complex plane, and the mode switching between throats will not take place.

In Section~\ref{sec:Tunnelling} we consider models that exhibit enhanced tunnelling rates, either due to resonant effects 
(such as the ones mentioned
in~\cite{henry, henry0602}) or because the shape of the potential barrier separating the
throats is modified (and the barrier is lowered). The basic idea is to
modify the throats near their top, where the effective potential barrier is largest. Examples we look at
include two throats with different curvature radii, resonant tunnelling through the gravity box
of~\cite{lykken}, and a toy 5-dimensional model representing two throats joined smoothly in the bulk, in
which the curvature is changing stepwise due to the presence of
additional 3-branes.\footnote{Our term ``3-branes'' in this sense simply means domain walls with arbitrarily
chosen tension; they are not supposed to represent string theory D3-branes.} We show in
Section~\ref{sec:Tunnelling} that low curvature in the central bulk 
region of the geometry can, under fairly general assumptions, lead to a drastic enhancement of tunnelling
rates for the low-lying KK modes. We end in Section~\ref{sec:Conclusions} with discussion and conclusions.

\mysection{Review: Model of a two-throat compactification}

\label{sec:background}

We start by reviewing the 5-dimensional model of~\cite{barnaby, dimopoulos0106}. We first summarize the
background geometry, then concentrate on the fluctuations of the 5-dimensional metric as in~\cite{lykken}.

\subsection{Background}

We start with an infinite Standard Model (SM) throat and in later sections generalize
our analysis to the case of a finite SM throat\footnote{According to a conjecture
of~\cite{rob}, during inflation the SM throat warping cannot be stronger than the square of the A
throat warping; it is assumed that after inflation ends the SM throat will settle into its ``vacuum'' (long)
states, though this process may not be finished at the time relevant for KK mode tunnelling and
decay.}.
Thus, our setup is Einstein's gravity in 5 dimensions parametrized by $(x^\mu, z)$ with $\mu=0,..,3$,
with a ``Planck'' (UV) 3-brane embedded at $z=0$ and an ``annihilation'' (A) brane embedded at $z=-z_A<0$. The
Planck and A branes have tensions $V$ and $-V$ $(V>0)$, respectively. The coordinate $z$ thus
runs from $-z_A$ to $\infty$. To simplify notation, however, we will use positive $z$ values on both sides of
the UV brane and will distinguish them by superscripts $A$ (annihilation side) and $S$ (Standard Model side). The
action is 
\beq
  S = \int \d^4x dz \sqrt{-g} \left(2M_5^3R - \Lambda \right) - \int d^4x \sqrt{-g_4(z=0)} V + \int
       d^4x \sqrt{-g_4(z=z_A)} V. \labell{action}
\eeq
At $z=z_A$ we impose, for the mode analysis, $\ZZ_2$ boundary conditions on both the background
{\em and} the fluctuations. In Section~\ref{sec:Tunnelling} we discuss the appropriate
boundary conditions for the tunnelling calculation.

The background consists of AdS bulk with metric
\beq
  ds^2 = \sigma^2(z) (dx^2 + dz^2), \labell{background}
\eeq
with the warp factor $\sigma(z)$ given by
\beq
  \sigma(z) = \frac{1}{1+k|z|}, \labell{sigma}
\eeq
and the curvature scale $k$ (equivalently, the radius of curvature $L\equiv 1/k$) is determined by the
cosmological constant $\Lambda >0$ as 
\beq
  k^2 = \frac{ -\Lambda}{24M_5^3}.
\eeq
As is well-known~\cite{RS1}, to get flat branes localized at definite
points in the $z$ direction, their tension $V$ must be tuned to
\beq
  V = \pm 24M_5^3k\, , \labell{BraneTension}
\eeq
where the plus and minus signs apply to the Planck and inflating branes, respectively. An extension of this
formalism to the case where the curvatures on both sides of the brane are not the same is discussed in
Section~\ref{ManyBraneRS}.

\subsection{Fluctuations}

\label{sec:Fluctuations}

We would like to study linearized fluctuations of the metric~\reef{background} along the lines
of~\cite{lykken}. We restrict ourselves to the metric fluctuations in the directions parallel to the branes
and parametrize the full metric as
\beq
  ds^2 = \sigma^2 \left[ (\eta_{\mu\nu} + \td h_{\mu\nu}) dx^\mu dx^\nu + dz^2 \right],
\eeq
with the fluctuations $h_{\mu\nu} = \sigma^2 \td h_{\mu\nu}$ satisfying the 4-dimensional transverse--traceless gauge conditions. 
Setting
\beq
  h_{\mu\nu} = e^{ipx} \sigma^{1/2} \psi_m(z)\epsilon_{\mu\nu}, \labell{hmunu}
\eeq
with $m^2 = -p^2$ being the four-dimensional mass of the fluctuations, 
the linearized Einstein equations for the metric fluctuations reduce to a one-dimensional problem
\beq
  \left[ -\half \prt_z^2 + V(z) \right] \psi_m(z) = \half m^2 \psi_m(z), \labell{PsiEom}
\eeq
with the potential $V(z)$ given by
\beq
   V(z) = \frac{15k^2}{8(k|z|+1)^2} - \frac{3k\sigma(z)}{2} \left(\delta(z) - \delta(z-z_A)\right).
              \labell{potential}
\eeq
The solution of~\reef{PsiEom} away from the branes is
\beq
  \psi_m(z) = N_m \, \left(m \tdz\right)^{1/2} \left[ Y_2(m\tdz) + Q_m J_2(m\tdz) \right],
              \labell{GeneralPsi}
\eeq
where we have introduced
\beq
  \tdz \equiv z + \frac{1}{k}  \labell{DefnOfTildeZ}
\eeq
to make the notation more compact.

We now take $\psi_m$ to be of the general form~\reef{GeneralPsi}, with coefficients $N_m^S, Q_m^S$
and $N_m^A$, $Q_m^A$ on the SM and A sides of the Planck brane, respectively. The coefficient
$Q_m^A$ will be determined by the jump condition coming from~\reef{PsiEom} at the A brane,
while $Q_m^S$ and the ratio $N_m^S/N_m^A$ will be determined by the jump condition and continuity at
the Planck brane. 

The following relation, derived using standard Bessel function identities and valid away from the
branes, will be useful:
\beq
  \frac{\prt_z \psi_m(z)}{\psi_m(z)} = -\frac{3k}{2} \sigma(z) + m\frac{Y_1(m\tdz) + Q_m
      J_1(m\tdz)}{Y_2(m\tdz) + Q_m J_2(m\tdz)}.
     \labell{DPsiOverPsi}
\eeq
Equation~\reef{PsiEom} and the $\ZZ_2$ conditions imposed on the fluctuations $h_{\mu\nu}$ at the
A brane lead to
\beq
  \prt_z \psi_m(z\ra z_A^-) = \frac{3k}{2} \sigma(z_A) \psi_m(z_A),
\eeq
giving 
\beq
  Q_m^A = -\frac{Y_1(m\tdz_A)}{J_1(m\tdz_A)}. \labell{QmI}
\eeq
Similarly,~\reef{PsiEom} at the Planck brane implies
\beq
  -\half \left( \prt_z \psi_m(0^+) - \prt_z \psi_m(0^-) \right) = \frac{3k}{2} \psi_m(0)
\eeq
leading to
\beq
  Q_m^S = - \frac{Y_1(m/k) \left[ Y_2(m/k) + Q_m^A J_2(m/k) \right] + Y_2(m/k) \left[ Y_1(m/k) + Q_m^A
                  J_1(m/k) \right] }
               {J_2(m/k) \left[ Y_1(m/k) + Q_m^A J_1(m/k) \right] + J_1(m/k) \left[ Y_2(m/k) + Q_m^A
                  J_2(m/k) \right] }. \labell{QmS}
\eeq
Since the SM side is semi-infinite, we have a continuum of plane-wave normalizable modes
with normalization constants given by~\cite{lykken}
\beq
  N_m^S = \sqrt{\frac{1}{1+(Q_m^S)^2}}. \labell{NmS}
\eeq
Imposing continuity of $\psi_m$ at $z=0$ implies
\beq
  N_m^A = N_m^S \frac{Y_2(m/k) + Q_m^S J_2(m/k)}{Y_2(m/k) + Q_m^A J_2(m/k)}.
         \labell{NmI}
\eeq

\subsection{Energy and curvature scales} 

\label{sec:Scales}

Before proceeding with our calculations, we should understand the physical scales in the
problem. In the RS scenario that serves as our toy model, we have several mass and length scales: the
5-dimensional Planck scale $M_5$, an independent scale
$k$ that sets the curvature radius of the AdS space and the tension of the branes,
the length $z_A$ of the A throat and an (apriori independent) scale $M_A$ of
the brane physics that produces the KK modes which are, in the warped reheating
scenario, responsible for the transfer of energy to the SM side. 

If our setup is to serve as a model of string theory compactifications, the parameters of our
model must be matched to the appropriate scales in string compactifications. Hence we 
would like to take $k$ to be somewhat below the strings scale $M_s$ 
\beq
  k = \frac{M_s}{\gamma g_s^{1/4}}
\eeq
with $\gamma$ a numerical factor of order say 10 that roughly corresponds to the compactification radius of
the internal manifold, and $g_s$ the string coupling. Parametrized this way, the factor $\gamma$ is related~\cite{gary, kofman} 
to the flux quantum numbers of the Klebanov--Tseytlin throat~\cite{KT, KS} as 
\beq
  \gamma \sim (MK)^{1/4} \, .
\eeq
The bare tension of the branes ($D3$ branes in string theory) is $\sim M_s^4/g_s$,
with $g_s<1$. The brane scale
$M_A$ would be given by the tension of the annihilating branes times the warp factor
at the position of the annihilating branes,
\beqa
  M_A &= &\frac{M_s}{g_s^{1/4}} \sigma_A \nn \\
      & = & \frac{\gamma}{g_s^{1/4}} k \sigma_A \nn  \\
      & \approx & \gamma \frac{1}{z_A} \, ,
              \labell{EffectiveBraneScale}
\eeqa
where we have denoted $\sigma_A \equiv \sigma(z_A)$. 
Thus, the effective mass scale on the A brane is somewhat (an order of
magnitude, say) larger than $1/z_A$. The 4- and 5-dimensional Planck scales $M_4, M_5$ are related to the string mass
$M_s$ as
\beq
  M_4^2 k = M_5^3 \sim M_s^3 g_s^{-3/4} \gamma^5 \, ,
\eeq
where we have assumed that the compactification volume of the internal 5 dimensions is $V_5 \sim k^{-5}$.
Expressed in terms of the string compactification parameters, the ratio $k/M_5$ is
\beq
  \frac{k}{M_5} \sim \gamma^{-8/3} \, . \labell{kM5RatioInStringPars}
\eeq  
The inflation scale $H$ is
different from $M_A$~\cite{rob}, and in this setup is 
\beqa
  H & \sim & \frac{1}{\gamma^3} M_4 \sigma_A^2 \nn \\
    & \sim & \gamma k \sigma_A^2 \, ,  \labell{HubbleScale}
\eeqa
\ie it differs from $M_A$ by an extra factor of roughly\footnote{Derivation of~\reef{HubbleScale}
again assumes $ V_5 \sim 1/k^5$.} 
$\sigma_A$. 

Assuming $H/M_4 \sim (10^{-8}$---$10^{-5})$ and the 4-dimensional Planck scale $M_4$ being not very
different\footnote{$M_4 = M_s \gamma^3/g_s$; with standard assumptions $M_4$ will be a few orders of
magnitude larger than $M_s$} from 
the string scale (and thus also from $k$) leads to the annihilation warp factor $\sigma_A \sim
(10^{-3}$---$1)$~\cite{rob}. To exhibit issues peculiar to warped inflation, in the remainder of this work we
will assume that $\sigma_A$ is in the low end of that range. 

\mysection{Two throats full of modes}

\label{sec:Modes}

In this section we study quantitatively the wavefunctions of KK modes likely to be produced in the decay of
closed string remnants of the brane annihilation. As opposed to~\cite{gary, kofman} who considered modes in
each throat separately, we study the modes of the full two-throat system. 

In Section~\ref{sec:Suppression} we revisit the enhancement of the KK mode wavefunctions compared to the
graviton deep in the A throat.  Most of our conclusions agree with the ones that~\cite{gary, barnaby,
kofman} arrived at. The calculations are quite technical, so we relegate the details to
Appendix~\ref{app:Suppression} and provide only a summary in the main text. We find that most of the modes of
the full two-throat system are suppressed at the A brane (compared to the graviton zero mode), 
even when their masses are of order of
$1/z_A$; only modes that fall within narrow ``resonance peaks'' of width 
\beq
   \delta m \sim m^5/k^4 \labell{PeakWidth}
\eeq
have their
wavefunctions enhanced. These narrow peaks correspond to the discrete spectrum of modes of the A
throat when the latter is viewed as a system on its own.

The mode analysis allows us to exhibit tunnelling between the throats 
as a wavepacket decoherence
effect in the combined system, which we describe in Section~\ref{sec:Decoherence}. Wavepackets
localized in the A throat will be composed predominantly of modes falling within the narrow resonance
peak; hence the tunnelling rate $\Gamma$ can be viewed as the rate at which the wavepacket decoheres and the destructive
interference in the SM throat disappears. The rate is given precisely by the (energy) width of the wavepacket,
\beq
  \Gamma = \delta m \sim \frac{m^5}{k^4} \, .
\eeq
This expression is analogous to the (inverse of the) ``throat interaction time scale'' given in~\cite{kofman}. 

In Section~\ref{sec:FiniteSMThroat} we consider modifications of the tunnelling rate that arise when one makes the
SM throat finite, with coordinate length $z_S$. The spectrum will be quantized roughly in units of $1/z_S$
(for $z_S \gg z_A$). We find that for very long SM throats the results are, as expected, unchanged. On the other hand,
when the SM throat becomes so short that the mode spacing $1/z_S$ becomes longer than
the width $\delta m$ of the resonant peaks~\reef{PeakWidth}, a mode localized in the
A throat will remain localized there forever (at the linearized level; interactions can lead to
decay of the mode). The surprising observation is that this suppression of tunnelling will occur when
\beq
  z_S < z_A \frac{k^4}{m^4} \, ,
\eeq
Interestingly, \cite{rob} argue that during inflation, the SM throat
cannot be longer than roughly $z_A k/m$ which is much shorter than the above bound; after inflation ends, the
SM throat presumable gradually relaxes to its full length, but the tunnelling may still be suppressed at the
time the KK modes decay. 

We now turn to detailed calculations.

\subsection{Deep down in the throat: KK modes vs. the graviton}

\label{sec:Suppression}

In any model of inflation it is important that after inflation ends, the fraction of energy that
ends up in gravitons be small.
In brane inflation this means that production of gravitons from brane
annihilation should be suppressed compared to other modes (in this case the KK modes) that can transfer the
energy released by annihilation to Standard Model matter.

The results of~\cite{gary} indicate that the flat-space
production rates of gravitons and KK modes in the decay of highly excited closed strings~\cite{llm} are roughly
(up to a factor of order 1) the same. Hence, in the warped case, the relative production rates 
of KK modes and gravitons will be determined predominantly by the size of their respective wavefunctions.
Thus, we should compare the wavefunctions of
low-lying KK modes ($m \ll k$) to the wavefunction of the zero mode. The KK modes were analyzed
in the previous section, while  
the zero mode, including its normalization, is%\footnote{Strictly speaking, this
%expression is valid only for $z_A = \infty$; it is a good approximation when
%$z_A \gg 1/k$, which is what we assume anyhow.}
\beq
  \psi_0 = \frac{\sqrt{k}}{(k\td z)^{3/2}}\, . \labell{ZeroModePsi}
\eeq
To quantify the relative magnitudes of the wavefunctions, we will study the ratio
\beq
  R_\psi(m) \equiv \frac{\psi_m(z_A)}{\psi_0(z_A)}\, , \labell{R_psi}
\eeq
and its physically meaningful integral (since the KK modes have a continuous spectrum)
\beq
  p(m) = \int_0^m dm' R_\psi^2(m')\, .  \labell{p_m}
\eeq
The function $p(m)$ can be interpreted as the ratio of probability densities to find a KK mode
of mass $m'<m$ and to find a graviton at the A brane. If the couplings of such KK
modes and the graviton to the closed string remnants of brane annihilation is similar, $p(m)$
will roughly represent the relative abundance of KK modes vs. gravitons in the products of decay of the
closed strings. Phenomenology requires that $p(m)$ be large.

One can consider KK modes with masses in two regimes: the first one will encompass ``very light'' KK modes 
in the sense that $mz_A \ll 1$. The second, ``intermediate'', regime contains masses such that $mz_A \sim 1$
while $mk \ll 1$. In a full two-throat system we have a (semi-)continuum of modes in both regimes; on the
other hand, an approximation with two separate, weakly coupled throats indicates that only a restricted subset
of modes should be important in the A throat, namely the modes of the full system that approximate
the modes of the A throat separated from the rest. This turns out to be the case.

% and along the 
%way we will find an alternate derivation of the time scale of tunnelling of the KK modes from
%the A throat into the SM one. 

Very light KK modes, $mz_A \ll 1$, are expected to be suppressed on the A side simply because the
A throat 
is shorter than the SM throat (which we assume to be infinite here). 
The mass is so small that the volcano potential~\reef{potential} at the
inflating brane is already large enough to suppress the mode: For $kz_A \gg 1$ the
potential is $V(z_A) \sim 1/z_A^2$, so one would expect that for the KK masses
$m<1/z_A$ the corresponding mode would be suppressed on the inflating brane. Explicit calculations in
Appendix~\ref{sec:SmallMassModes} bear this expectation out: we find 
\beq
  R_\psi^2 \approx \frac{m}{k^2}
\eeq
and
\beq
  p(m) \approx \half \frac{m^2}{k^2} \ll 1\, .
\eeq
Hence, production of such low-lying KK modes will be greatly suppressed compared to the production of
gravitons.

The situation is more interesting in the intermediate regime $mz_A \sim 1$. In
Appendix~\ref{sec:MediumMassModes} we find that for generic masses, the corresponding KK modes are still
suppressed,
\beq
   R_\psi^2  \sim \frac{m}{k^2} \, ,
\eeq
so
\beq
  p(m) \sim  \frac{m^2}{k^2} \sim \sigma_A^2 \, .
\eeq
Here the equality holds only up to factors of order 1, which we denote by the $\sim$ symbol. This suppression
is counterintuitive: one would expect the Bessel functions entering the KK mode wavefunctions to begin
approaching their asymptotic regime and hence be enhanced (in the integrated ratio $p(m)$) by a factor of
roughly $\sigma_A^{-2}$. To resolve this discrepancy, it helps to look at the plot of $R_\psi$ as a function
of the mass $m$, Fig.~\ref{fig:RatioOfPsi2}(b). The plot shows that modes with ``generic'' mass are, in
fact, suppressed as indicated above; however, the plot also shows narrow resonant spikes in which the ratio
$R_\psi$ is large. The width of the spikes is estimated in Appendix~\ref{app:Suppression} and turns out to be
given by the formula~\reef{PeakWidth},
\beq
  \delta m \sim m^5/k^4 \nn \, .
\eeq
The contribution of the spikes to the integrated ratio $p(m)$ turns out to be large and in fact
confirms the intuitive argument for the KK mode enhancement:
\beq
  p(m)_{spike} \sim \sigma_A^{-2} \, .
\eeq
This resonance phenomenon may seem surprising, but in fact it is to be expected in a system that is
effectively composed of two weakly coupled potential wells. Those modes of the full system that have a small
overlap with the ``native'' modes of the A throat (viewed as a separate system) will be suppressed
there; the only modes of the full system that can be large are those that have a large overlap with the
native modes of the A throat.  

%
%We now turn to detailed calculations underlying the above results.
%

\subsection{Tunnelling as a decoherence effect}

\label{sec:Decoherence}

Having modes of the full system allows us to exhibit the tunnelling of KK modes from the
A throat to the SM one as a decoherence phenomenon: creation of a KK mode in the A throat is
viewed as setting up a wave packet localized near the bottom of the
A throat. As usual in quantum mechanics, to study the time evolution of the packet one
can, \eg
expand it into eigenmodes of the full system (whose time evolution is known precisely). The initial
expansion into eigenmodes is quite special (\ie coherent), giving zero throughout most of the full system
(namely in the SM throat). As the eigenmodes evolve with different frequencies, the evolution will destroy
the coherence and the wavepacket's wavefunction will develop a nonzero generic amplitude in the SM throat,
corresponding to the KK mode tunnelling out of the A throat. Let us call the decoherence time, \ie
the time it takes the (initially highly correlated) expansion coefficients to become essentially random,
$t_{tun}$. One can estimate $t_{tun}$ as the inverse of the spread in frequency, $(\delta m)^{-1}$, of the initial
wave packet, or the decoherence (tunnelling) rate $\Gamma_{tun}$ as the spread itself.
 As we have argued in the previous subsection, the KK modes in the A throat will be
predominantly produced in the narrow bands around the peaks of $R_\psi$; the width of these peaks was
estimated in~\reef{deltam} and we have
\beq
  t_{tun}^{-1} = \Gamma_{tun} \sim \delta m \sim \frac{m^4}{k^4} \frac{1}{\td z_A} \sim \frac{m^4}{k^4} m \, .
             \labell{PlaneWaveTunnellingRate}
\eeq
The same expression was derived in Section 3 of~\cite{gary}, and a similar one was given Section 5 of~\cite{kofman} as
the ``interaction time scale''; however, the
authors of~\cite{kofman} assumed the UV scale to be the 4-dimensional Planck mass $M_4$; our calculation
above (and the calculations of~\cite{dimopoulos0106}) confirm that the relevant UV scale here is the
curvature scale $k$ which is expected to be somewhat smaller than $M_4$.

\subsection{(Non-)Tunnelling into a finite SM throat}

\label{sec:FiniteSMThroat}

In deriving~\reef{PlaneWaveTunnellingRate} we have assumed that the SM throat is infinite, even though in reality it will be
long but finite; we will now clarify how long it must be to be for our preceding calculations to apply
(inasmuch as our results agree with the ones in~\cite{gary, kofman}, this restriction applies to those works
as well). Hence, in this section we will assume the SM throat is cut off at $z=z_S > z_A$, with the $>$
sign meaning at least 1 order of magnitude.

If the SM throat is finite with length $z_S$, KK modes of the full system will be quantized roughly in units
of $\delta_s m \sim 1/z_S$. One might think that simply having $z_S > z_A$ would mean that from the
A throat point of view we a have a quasi-continuous spectrum, but this is not the case. The
fastest way to see it is to consider the resonant spikes in $R_\psi$ discussed in
Section~\ref{sec:Suppression} and Appendix~\ref{sec:MediumMassModes}: the width of these spikes is $\delta m \sim (m/k)^4 z_A^{-1} \ll
z_A^{-1}$. In fact, if the spacing $\delta_s m$ is larger than the width of the spikes, it is likely that
none of the modes present in the system will fall within the interval $\delta m$, indicating a suppression
of tunnelling. This argument is a bit naive in that in addition to the $\delta_s m$ spaced modes
(that one can consider to be native to the SM throat) there will be extra modes with spacing $\delta_A m \sim
1/z_A$ that are mostly localized in the A throat; however, precisely because these modes are
localized in the A throat, wavepackets consisting mainly of such modes cannot tunnel.

To put some firm math behind the intuition we consider modes in the finite two-throat system explicitly. The
boundary condition~\reef{QmI} now applies to the SM side as well (with the obvious modification),
\beq
  Q_m^S = -\frac{Y_1(m\tdz_S)}{J_1(m\tdz_S)} \, . \labell{QmSCut} 
\eeq
Modes of the full system have masses $m$ for which $Q_m^A$ and $Q_m^S$ satisfy~\reef{QmS}. 

The discrete spectrum of modes implies a change in normalization: just like the zero mode, the massive modes
are now normalizable in the conventional sense,
\beq
  \int dz\, \psi_m^2(z) = 1 \, . \labell{NewNormalization}
\eeq
Fortunately, the normalization integrals for the wavefunctions given by~\reef{GeneralPsi} can be evaluated in
a closed form. We list them here for future use and define a shorthand notation: 
\beq
  C_J(x) \equiv \int dx\, x J_2^2(x) = \half x^2 \left[ J_2^2(x) - J_1(x)J_3(x) \right] \, , \nn \\
\eeq
\beq
  C_Y(x) \equiv \int dx\, x Y_2^2(x) =\half x^2 \left[ Y_2^2(x) - Y_1(x)Y_3(x) \right] \, , \nn \\
\eeq
\beq
  C_{JY}(x) \equiv \int dx\, x Y_2(x)J_2(x) = \frac{1}{4} x^2 \left[ 2J_2(x)Y_2(x) - J_1(x)Y_3(x) 
                          - Y_1(x) J_3(x) \right] 
         \, .  \label{NormalizationIntegrals}
\eeq
To get oriented, let us first look at modes that are mostly localized in the SM throat (these are the
equivalents of the ``generic'' modes discussed in Appendix~\ref{sec:MediumMassModes}). Such modes have $Q_m^A
\sim 1$ (\ie the Bessel functions entering~\reef{QmI} have generic values); this implies
\beq
  Q_m^S \sim \frac{k^2}{m^2} \gg 1 \, ,
\eeq 
meaning that the masses of such modes lie near the zeros of $J_1(m\tdz_S)$, confirming the intuition that the
mass would be quantized in units of $1/z_S$. Since $Q_m^S \gg Q_m^A$, the normalization will be determined
predominantly by $C_J(m\tdz_S)$ and works out to be
\beq
  N_m^S \sim \frac{m}{Q_m^S C_J(m\tdz_S)} \sim \frac{m^2}{k^2 \sqrt{z_S}} \, .
\eeq
Continuity of $\psi_m$, Eq.~\reef{NmI}, then gives
\beq
  N_m^A \approx N_m^S \, ,
\eeq
and for the wavefunction itself we find
\beq
  \psi_m^A(z_A) \sim \frac{m^2}{k^2 \sqrt{z_S}} \, .
\eeq
Comparing this with the zero mode wave function~\reef{ZeroModePsi} we find
\beq
  R_\psi^2 (m) \sim \frac{m^4 \tdz_A^3}{k^2 \tdz_S} \sim \frac{m \, \delta_s m }{k^2} \, .
\eeq
Summing over all modes whose $m\sim z_A$ we find the equivalent of~\reef{PForMediumMass},
\beq
  p(m) = \sum_{m' \leq m} R_\psi^2(m') \sim \frac{m^2}{k^2} \, .
\eeq
(We have not explicitly considered the very light modes, but just as in the continuous case, their
contribution scales the same with $m/k$.)
This means that the generic (or SM throat) modes are small in the A throat. In the continuum case we
found that most of the KK mode presence was due to a narrow range of modes that we called resonant. In the
discrete case such modes occur when $Q_m^S$ is (nearly) zero and large $Q_m^A$ is large: $Q_m^S = 0$
implies, via~\reef{QmS}, 
\beq
  Q_m^A \sim \frac{k^2}{m^2} \gg 1 \, .
\eeq 
In this case the biggest contribution to the normalization integrals can come either from $C_Y(\tdz_S) \sim
m\tdz_S$ or from $C_J(\tdz_A) \sim m\tdz_A k^4/m^4$, depending on the relative size of $z_A, z_S$ and $1/k$. 

If the SM side dominates, namely for
\beq
  z_S > z_A \frac{k^4}{m^4} \, , \labell{SMSideDominates}
\eeq
the mode will be mostly localized on the SM side despite being large in the A throat, simply because
the SM throat is so long that the probability of finding the particle in it is close to 1. This is the case
that closely approximates the continuum setup: the resonant KK modes are large in the A throat, yet
if a wave packet localized in the A throat is formed from them, it will tunnel to the SM throat as
discussed in Section~\ref{sec:Decoherence}. Eq.~\reef{SMSideDominates} confirms the intuitive arguments
given at the start of this section: precisely when the former is satisfied, the spacing of the SM throat
modes is fine enough that some will fit into each resonant spike. 

On the other hand, if the A side dominates, namely for 
\beq
  z_S < z_A \frac{k^4}{m^4} \, , \labell{ASideDominates}
\eeq
we find a surprising result: the KK modes that are large in the A throat have a very small
probability of being detected in the SM throat, and if a wave packet of such modes is formed in the
A throat, it will generically remain there\footnote{It may be possible to form a coherent wave
packet that will appear very briefly in the SM throat, but it will decohere quickly and the particle(s) will
reappear in the A throat again.}; only a small part of order $z_S m^4/(z_A k^4) < 1$ of these KK modes will
tunnel. Note that this can be the case even if the A throat is much shorter: \eg if $m\sim 1/z_A
\sim 10^{-3}k$, the A throat can be almost 12 orders of magnitude shorter!  

The preceding arguments provide a further constraint on the relative length of the throats: since the modes
relevant to tunnelling have masses of order $1/z_A$, Eq.~\reef{SMSideDominates} implies
\beq
 z_S > \sigma_A^4 z_A \, . \labell{zSConstraint}
\eeq
Written in terms of warp factors, we have a constraint on the Standard Model warp factor $\sigma_S \equiv
\sigma(z_S)$ that reads
\beq
  \sigma_S < \sigma_A^5 \, . \labell{SigmaSConstraint}
\eeq
This is a stringent constraint; in fact, \cite{rob} suggests that during and shortly after
inflation, the warp factor of the SM throat is no less than the square of the A warp factor, 
\beq
  \sigma_S \gsim \sigma_A^2 \, ,
\eeq
and hence the inequality~\reef{SigmaSConstraint} would not be satisfied.\footnote{As we explain in
Section~\ref{sec:DifferentCurvatures}, it is possible, if somewhat contrived, to
relax~\reef{SigmaSConstraint} if the throats have different curvatures.}

In Appendix~\ref{app:DeltaBarrier} we show that a condition similar to~\reef{SMSideDominates}
can be derived in a toy model of a $\delta$-function potential barrier separating two finite flat boxes. 
A natural question to ask is what is the physical interpretation of the factor $m^4/k^4$ relating the two
throat lengths in~\reef{SMSideDominates} and its analog in the $\delta$-barrier case. We propose that, in
general, tunnelling will occur only if the level spacing of the target side is finer than the level spacing
on the source side multiplied by what one may call ``plane wave tunnelling probability'' $P$,
\beq
  (\delta m)_{target} < P (\delta m)_{source}  \, . \labell{GeneralTunnellingCondn}
\eeq
The tunnelling probability $P$ is the ``usual textbook'' probability of a particle that starts as a plane wave
far from the barrier on the source side to emerge as the corresponding plane wave on the target side. For the
Randall--Sundrum model $P$ was calculated by~\cite{dimopoulos0106} and found to be $P \sim m^4/k^4$; we show
in Appendix~\ref{app:DeltaBarrier} that~\reef{GeneralTunnellingCondn} also applies to the $\delta$ potential
barrier.

In retrospect, at least naively, this looks like an
obvious statement: the size of the wavefunction squared, \ie the probability density, in the target well
is suppressed by the tunnelling rate with respect to the probability density in the source well; hence
their lengths must differ by the corresponding factor for the probabilities (which are integrals of the
densities) to be comparable. However, the statement uses the tunnelling rate of plane waves (which is
independent of the lengths of the wells) to compare the tunnelling probabilities of wave packets made of
standing waves in finite wells (and in finite wells the spectrum and tunnelling probabilities do depend on
the lengths of the wells), so perhaps the result is not so obvious after all.

Incidentally, the above result, if true in general, would be applicable also to the multibrane setups considered in
Section~\ref{sec:Tunnelling}; the enhancement of the tunnelling rates of plane waves implies that the ratio
of throat warp factors need not be as large as indicated in~\reef{SigmaSConstraint}.

Before we end this section, a comment is in order about an apparent discrepancy with~\cite{gary}, whose
authors claim that tunnelling 
rates are essentially (\ie up to factors of order 1) unaffected by making the SM throat short. The calculation
in Appendix A of~\cite{gary} appears to agree with ours above (although it is formulated in different terms),
but their interpretation is incorrect in the sense that~\cite{gary} split standing waves into in- and out-going
parts and then regard the latter as independent, interpreting the outgoing component as representing a flux
of tunnelled particles. 

\mysection{Energy transfer via decay without tunnelling}

\label{sec:Decay}

If the throat lengths are such that A throat KK modes remain localized in the A throat
indefinitely, the said KK modes will decay via the nonlinear interactions present in the
Einstein--Hilbert Lagrangian (and via the coupling of the 5-dimensional gravitational fluctuations to
the SM brane fields). One might expect that for the lowest A throat KK mode
the only decay channel would be pairwise annihilation into gravitons,
but this is not so: the wavefunction of A throat KK modes has a small tail in the SM throat, and the tail 
actually causes a sizable decay rate into SM throat KK modes. In this section we study this decay rate and
find that in general it appears to be faster than the plane wave tunnelling rate. 

Before we dive into the calculations, however, a cautionary note is in oder. It
is clear that the amplitude of the A throat KK wavefunction tail in the SM throat depends to a large degree
on the potential shape in the UV region. For example, Ref.~\cite{henry0602} 
found that tunnelling out of a full AdS$_5$ throat into the bulk is suppressed compared to the analogous
quantity in the RS model; this suppression can be viewed as due to a modification of the effective potential
entering the Schr\"odinger Equation~\reef{PsiEom}. While the
Randall--Sundrum geometry is a good approximation deep in the throat, the region near the UV brane will
presumably look very different in (a dimensional reduction of) a full string compactification. With this
possible loophole in mind, let us proceed with the actual calculation of the decay rates.

We start by quantifying the interaction strengths, following~\cite{barnaby}. 
The 5-dimensional Einstein--Hilbert Lagrangian $M_5^3 \sqrt{\td g}R$ gives rise to an infinite series of
interactions of the (canonically normalized) fluctuations $h$ of the metric $\td g$ around its background value $g$
($\td g= g + M_5^{3/2}h$). Schematically they are of the form 
\beq  
 \cL_n \sim \frac{1}{M_5^{3n/2-3}} \int \sqrt{g} \, \nabla^2 \left[\left(g^{-1}\right)^{n+1} h^n
                \right]  \labell{InteractionLagrangian}
\eeq
where the derivatives can act both on the background metric and the fluctuations. We will only be
interested in the powers of the warp factor and will neglect all order 1 coefficients (in particular,
we will not attempt to keep track of the factors arising from various possibilities of contracting
the indices in~\reef{InteractionLagrangian}). 

A $z$-derivative acting on $g^{-1}$ will give $k\sigma g^{-1}$ (derivatives in the $x^\mu$
directions are zero). A $z$-derivative acting on $h$ will produce several terms that can be roughly
characterized as being either of the form $k\sigma h$ or $m h$, where $m$ is the mass of the KK
mode. Derivatives of $h$ in the $x^\mu$ directions will simply multiply the wave function by the
corresponding momentum. Hence the interaction~\reef{InteractionLagrangian} 
leads to the following 3 basic types of terms,
\beq
  \cL_n \sim \frac{1}{M_5^{3n/2-3}}  \int \sqrt{g} \left(g^{-1}\right)^{n+1} h^n 
   \left\{ (p\cdot p, mm), mk\sigma, (k\sigma)^2 \right\} \labell{nPtCouplings}
\eeq
(We consider the $p\cdot p, mm$ terms to be of the same type; note that the $p\cdot p$ and $mm$ symbols
should not be taken at face value; they could mean, \eg $p_1 \cdot p_2$, $m_1m_2$ if the derivatives act on
different KK modes.) The terms differ by their $z$ dependence and dependence on momenta.   

To obtain an effective 4-dimensional interpretation, we perform the integral over $z$
in~\reef{nPtCouplings}; the results will be an effective 4-dimensional Lagrangian. We will treat
all terms in~\reef{nPtCouplings} together; denote the power of $k\sigma$ by $\alpha$ (so
$\alpha=0,1,2$); then we can summarize~\reef{nPtCouplings} as
\beq
  \cL_n \sim \frac{1}{M_5^{3n/2-3}}  \int d^4x dz \,  \sqrt{g} \left(g^{-1}\right)^{n+1} h_1\ldots h_n
w_{(\alpha)} k^\alpha \sigma^\alpha
      \labell{nPtCouplings2}
\eeq
where $w_{(0)}$ is a combination of $mm$ and $p\cdot p$, $w_{(1)} = km$ and $w_{(2)} = k^2$.

We recall from Section~\ref{sec:Fluctuations} that the $z$-dependence of the graviton fluctuations
$h$ is given by the KK mode wave function and the appropriate prefactor (we neglect the Lorentz
tensor structure),
\beq
  h(x,z) = \sigma^{1/2}(z) \psi(z) \phi(x)
\eeq
where $\phi(x)$ is the 4-dimensional wavefunction that is usually taken to be simply $\exp(ipx)$. 
The $n$-point couplings can then be written concisely as
\beq
  \cL_n = \int d^4 x \, \zeta_n  \phi_1(x) \phi_2(x) \phi_3(x) \, ,
\eeq
where the ``coupling constant'' $\zeta_n$ (which in fact can contain factors of momenta) is the overlap integral along
the $z$ coordinate, 
\beq
  \zeta_n \sim \frac{1}{M_5^{3n/2-3}} \int dz \, \sqrt{g} \left(g^{-1}\right)^{n+1} \sigma^{n/2} w_{(\alpha)} k^\alpha
\sigma^\alpha \psi_1(z)\ldots \psi_n(z) \, . \labell{zetan}
\eeq
Amplitudes involving at least one graviton deserve a special consideration.\footnote{We thank Henry Tye,
Xingang Chen and Irina Mocioiu for corrections, discussions and suggestions on this point.} From the
perspective of the effective 4-dimensional theory, (4-dimensional) general
covariance requires that the graviton couple to the (4-dimensional) stress tensor, which must in this case be
proportional to (4-dimensional) momenta; hence amplitudes that include a graviton can only be nonzero for
$\alpha=0$.

Evaluating the integral~\reef{zetan} is in principle
(numerically) straightforward, but giving meaningful estimates is somewhat tedious because the
dominant regions in the $z$-integrals depend on how many A throat KK modes, SM throat KK modes and
gravitons enter the amplitude. Hence we will calculate the 3-point coupling~\reef{zetan} for 1
A throat KK mode with mass $m_0$ decaying into two SM throat KK modes with masses $m_{1,2}$. 

In principle one could contemplate the coupling of the A throat KK mode to a graviton and an SM throat KK mode, or to two
gravitons, but these couplings vanish: because of the graviton profile, a 3-point amplitude with $\alpha=0$
and one graviton reduces to the scalar product (of the z-dependent wavefunctions) of the two remaining modes
in the amplitude, which is zero whenever the two modes are different. (As noted earlier, the $\alpha\neq 0$
couplings involving gravitons must vanish by 4-dimensional general covariance.)

To perform the integration, we approximate the Bessel functions by their small argument
approximations and asymptotic forms respectively, where appropriate. In the region of $\tdz
<1/\sqrt{mk}$, which we will call the UV regime because it describes the region near the UV brane, the
KK mode wavefunctions have the same functional form as the graviton wavefunction, but their
normalization constants differ. For integrals that are mainly localized in the UV region we can neglect the oscillatory
nature of the Bessel functions at large arguments; for integrals that receive large contributions from the asymptotic
region, we will have to take the oscillations into account.

Near the UV brane in the A throat ($\tdz < 1/\sqrt{mk}$), the wavefunction of an A throat KK
mode behaves roughly as
\beq
  \psi(z) \sim \sqrt{m} \sigma_A^{1/2} \sigma^{3/2}(z) \, , \labell{ApproxIKKPsiIUV}
\eeq
while far from the UV brane ($m\td z \sim 1$) it is 
\beq
  \psi(z) \sim \frac{1}{\sqrt{z_A}} = \sqrt{k} \sigma_A^{1/2} = \const \, .
\labell{ApproxIKKPsiIIR}
\eeq
In the asymptotic region of the SM throat, the wavefunction of an A throat KK mode is
\beq
  \psi(z) \sim \frac{1}{\sqrt{z_A}} \frac{m^2}{k^2} = \sqrt{k} \sigma_A^{1/2} \frac{m^2}{k^2} =
         \const \, .  \labell{ApproxIKKPsiSMIR}
\eeq
The wavefunction of an SM throat KK mode with mass $m_i$ in the asymptotic region of the SM throat is 
\beq
  \psi(z) \sim \frac{1}{\sqrt{z_S}} = \sqrt{k} \sigma_S^{1/2} = \const \, ,
\labell{ApproxSMKKPsiSMIR}
\eeq
while in the A throat it is (here we assume $m_i \tdz_A <1$, so only the UV regime is
realized)
\beq
  \psi(z) \sim \frac{m_i^2}{k^2\sqrt{z_S}} \sqrt{m_i\tdz} \frac{1}{(m_i\tdz)^2} = \sqrt{m_i} \sigma_S^{1/2}
\sigma^{3/2}(z) \, .
\labell{ApproxSMKKPsiIIR}
\eeq
As we remarked above, the functional form of the KK modes in the UV regime and the graviton (given
by~\reef{ZeroModePsi}) is the same, only the normalization constants differ.

Under the approximations~\reef{ApproxIKKPsiIUV}--\reef{ApproxSMKKPsiIIR} the integrand
in~\reef{nPtCouplings2} will be a certain power of the warp factor $\sigma(z)$ (multiplied by various
constant factors). If this power is
negative or zero throughout most of the throat in which the integral is performed, the IR region of
the throat will dominate the integral; however, if the power of $\sigma(z)$ is positive, the UV
region will dominate. 

The amplitude of interest here is the decay of an A throat KK mode. We will assume that we are dealing with the
lowest-lying A throat mode; higher modes either decay into the lowest one or decay via the
same channels as the one we are considering. The coupling to two SM throat KK modes
is dominated by the IR region of the SM throat and is largest for the derivative coupling ($\alpha=0$). The result is
\beq
  \zeta^S_{A,2S} \sim \xi \frac{k^{1/2}}{M_5^{3/2}} w_0 \frac{m_0^2}{k^2}  \sigma_A^{1/2}
     \sigma_S^{-3/2} \, . \labell{CplS_A2S}
\eeq
In this case, the oscillatory nature of the wavefunctions cannot be neglected since the
integrand is large in the asymptotic region. To remind us of the fact that the naive expression is an
overestimate, we have inserted a factor of $\xi$ in~\reef{CplS_A2S} to account for the extra
suppression. The factor $\xi$ can be roughly estimated as follows. The integrand is $\sim
\sigma^{-3/2}(z)$ times a factor of type $\cos(m_i \tdz)$ for each wavefunction. The naive estimate of the
integral, integrating $z$ from $0$ to $z_S$, would be $\sim \sigma_S^{-5/2}/k$. An improved estimate
can be made by noting that $\prod \cos(m_i\tdz)$ can be written as a sum of sines and cosines of
various sums and differences of the arguments $m_i\tdz$. Hence a better approximation to the
integrand would be $\sigma^{-3/2}(z) \cos(m\td z)$; for a generic value of $m$ the leading term in
the integral would go as
$\int dz \sigma^{-3/2}(z) \cos(m\td z) \sim \sigma_S^{-3/2}/m$. Comparing this result with the naive
value fixes the correction factor to be $\xi \sim 1/(m\tdz_S) \sim \sigma_S/\sigma_A$. We note
that if the masses $m_i$ entering the amplitude are tuned such that (say) $m_0 = m_1 + m_2$, the
naive estimate is actually correct (as can be seen by taking a limit $m\ra 0$ of the improved
integral). However, this ``resonance'' cannot be used to enhance the coupling, because it happens
precisely when the decay products are on the threshold and have zero available phase space.
Including the correction factor gives an improved estimate of the $\zeta^S_{A,2S}$ coupling,
\beq
  \zeta^S_{A,2S} \sim \frac{k^{1/2}}{M_5^{3/2}} w_0 \frac{m_0^2}{k^2}  \sigma_A^{-1/2}
    \sigma_S^{-1/2} \, . \labell{CplS_A2S_Amp}
\eeq
The couplings can be written in various forms using $m_0/k \sim \sigma_A$ and defining the Standard
model mass scale $M_{SM} = M_5 \sigma_S$; the ratio $k/M_5$ (an independent parameter in the
RS model) is in a string compactification determined by the compactification moduli.

To obtain decay rates from the above couplings, one must perform a sum over accessible final states.
In Appendix~\ref{app:DecayRates} we perform the sums in detail; however, up to numerical factors, the
correct results can be obtained by simply assuming that all momentum factors are of order $m_0$ and
simply multiplying the resulting coupling (or decay rate) by the number of accessible modes, which in
case of a decay into two KK modes it will be $n_{KK}^2$.  
For the decay into two SM throat KK modes, the coupling~\reef{CplS_A2S} and~\reef{GammaKKFinal} give
the decay rate 
\beq
  \Gamma_{A,2S} \sim  k \left(\frac{k}{M_5}\right)^3 \sigma_A^{8} \sigma_S^{-3} \, . \labell{GammaA2S}
\eeq
This rate is quite fast; in particular, by making the SM throat long enough, it could be made much faster than the
plane wave tunnelling rate~\reef{PlaneWaveTunnellingRate}. However, in the next section we show that the actual decay
rate is slower than~\reef{GammaA2S}.

\mysection{Tunnelling is back: Effects of decay on KK mode wavefunctions}

\label{sec:DecayEffects}

In the previous section we have found the decay rate~\reef{GammaA2S} of an A throat KK mode in the SM throat. The A
throat KK modes can also decay in the A throat itself by pairwise annihilation into gravitons; the rate of this process
was estimated in~\cite{gary} to be
\beq
  \Gamma_{A,2g} \sim k \sigma_A^3 \, . \labell{GammaA2g}
\eeq
Generically, the throat lengths are expected to be such that $\Gamma_{A,2S} \gg \Gamma_{A,2g}$.
In this section we would like to incorporate the effects of unequal decay rates into our
treatment of the two-throat system.\footnote{We were motivated by 
Kofman and Yi~\cite{kofman} who suggest, by treating the two-throat system as two separate
throats coupled weakly through the barrier, that differing decay rates (represented as differing
imaginary parts in the modes' frequencies) will suppress tunnelling rates.}

Let us first use intuitive arguments to show what we expect to find. Wavefunction decay can be modelled by
the presence of a sink (an imaginary term in the potential). If the sink varies as a function of the spatial coordinate
(in our case, $z$), the 
wavefunction will tend to decay at different rates in different regions. Such a process would increase its
gradient energy too much; to counteract it, there will be a non-zero probability current from regions with a
small sink to regions with a large sink. In the case of two throats with different (but constant within
each throat) sinks we may expect that the throat with the smaller sink will act as a ``reservoir'' from which
the probability density will slowly ``seep'' into the throat with the big sink. The seepage between throats 
is expected to be slow because of the presence of the potential barrier. In the throat with the big sink, we have a
source (incoming current density) at the top of the throat while the sink is present throughout, so we would
expect the wavefunction to be largest near the source and decrease as a function of the coordinate $z$ along
the throat.  A toy model of such a situation for a $\delta$-function barrier separating two square wells 
is presented in Appendix~\ref{app:DeltaBarrierWithSink}; the model confirms the intuition outlined above.

An additional feature of the RS setup is that the decay rates, and hence also the sinks, depend on wavefunction
overlaps; as we argued above, the wavefunction of the decaying mode will, in turn, depend on the decay rates. Thus a
would-be large decay rate in the SM throat would mean that the overlap integrals are actually small, so the actual
decay rate cannot be too fast. How fast is too fast?
Physically, the decay rate in the SM throat cannot be faster than the rate at which probability
density leaks from the A throat into the SM throat -- the presence of the barrier here turns out to be the limiting
factor in the decay rate. We show that in the presence of a large sink, the wavefunction in the SM throat actually
becomes a (spatially decaying) outgoing wave, and the ``leakage rate'' becomes simply the plane wave tunnelling rate
through the barrier.  

We now describe detailed calculations supporting the intuitive arguments just given. We start by analyzing
the wavefunctions of the KK modes in the presence of sinks and then give an improved calculation of the decay
rate $\Gamma_{A, 2S}$ from Section~\ref{sec:Decay}.

\subsection{KK modes in the presence of sinks}

\label{sec:ModesWithSink}

The effect of decay of KK modes will be represented by an imaginary term $iS$ in the potential $V(z)$ in the
effective Schr\"odinger equation~\reef{PsiEom}.
We will take $S$ to be constant in each throat, but the values $S^A$ and $S^S$ in the A and SM
throats, respectively, will be different (this is a simplification of the real problem where the decay
term varies smoothly with $z$; the advantage is that it has a simple analytic solution). The
5-dimensional graviton fluctuations still have the form~\reef{hmunu}, but now we allow the frequency $ p^0 = \omega = m +
i\Gamma$ to be complex. The effective Schr\"odinger equation is then
\beq
  \left[ -\half \prt_z^2 + V(z) + iS \right] \psi_m(z) = \half \omega^2 \psi_m(z), \labell{PsiEomWithSink}
\eeq
The solution is a generalization of~\reef{GeneralPsi}: in the A and SM throats we have,
respectively,
\beq
  \psi_m(z) = \left \{ 
   \begin{array}{lc}
        N_m^A \sqrt{k\tdz} \left[ Y_2(\lambda^A\tdz) + Q_m^A J_2(\lambda^A\tdz) \right] \ \ \ &
\mbox{for} \ z<0\, , \\
  & \\
        N_m^S \sqrt{k\tdz} \left[ Y_2(\lambda^S\tdz) + Q_m^S J_2(\lambda^S\tdz)  \right]\ \ \ &
\mbox{for} \ z>0\, ,
   \end{array} \right. 
        \labell{GeneralPsiWithSink}
\eeq
where the (complex) wavenumbers $\lambda^A, \lambda^S$ are now different from the frequency (and from each
other),
\beqa
  \lambda^A &=& \sqrt{ \omega^2 - 2iS^A} \, , \labell{RSlambdaA} \\
  \lambda^S &=& \sqrt{ \omega^2 - 2iS^S} \, . \labell{RSlambdaS}
\eeqa
The matching conditions at the ends of the throats and at the UV brane are direct generalizations of
the ones derived in Section~\ref{sec:Fluctuations}: at the ends we have
\beqa
  Q_m^A &=& -\frac{Y_1(\lambda^A\tdz_A)}{J_1(\lambda^A\tdz_A)} \, , \labell{QmIWithSink} \\
  Q_m^S &=& -\frac{Y_1(\lambda^S\tdz_S)}{J_1(\lambda^S\tdz_S)} \, , \labell{QmSWithSink}
\eeqa
while the jump and continuity conditions at the UV brane can be written as
\beqa
  \lambda^A \frac{Y_1(\lambda^A/k) + Q_m^A J_1(\lambda^A/k)}{Y_2(\lambda^A/k) + Q_m^A J_2(\lambda^A/k)}
    &=& -  \lambda^S \frac{Y_1(\lambda^S/k) + Q_m^S J_1(\lambda^S/k)}{Y_2(\lambda^S/k) + Q_m^S
        J_2(\lambda^S/k)} \, ,  \labell{RSJumpWithSink} \\
  N_m^A \left[ Y_2(\lambda^A/k) + Q_m^A J_2(\lambda^A/k) \right] &=& 
  N_m^S \left[ Y_2(\lambda^S/k) + Q_m^S J_2(\lambda^S/k) \right] \, . \labell{RSContinuityWithSink}
\eeqa
Let us see how the spectrum will look like. We will assume (based on the decay rate calculations) 
that the magnitude of the sink terms is much smaller than the curvature scale $k^2$ (indeed if they were
not, bulk effects would invalidate our effective field theory analysis), though the sink terms can
be comparable in magnitude to the typical (squared) KK mode mass scales $1/z_A^2$. Then we have
$|\lambda^{A,S}| \ll k$ and we can apply the small-argument approximations~\reef{SmallArgument} to
the Bessel functions in the UV brane jump condition~\reef{RSJumpWithSink}. Analogously to the
sink-free case, the jump condition can only be satisfied when one of the $Q_m$ coefficients is large.
For definiteness let us look at the case when $Q_m^A$ becomes large (such modes will again be the
A throat modes). From~\reef{QmIWithSink} we find that $J_1(\lambda^A\tdz_A)$ must be small, so
$\lambda^A\tdz_A$ must be near one of the roots of $J_1$, all of which are real. Hence $\lambda^A$
will be (nearly) real (it will have a small imaginary part that is not important for us). This fixes
$\lambda_S$ to be
\beq
  \lambda_S = \sqrt{ (\lambda^A)^2 - 2i(S^S-S^A) } \, , \labell{LambdaSFromLambdaA}
\eeq
so $\lambda_S \equiv a+ib$ will have a substantial imaginary part $b$. The most important consequence
of $b$ is that in the asymptotic region of the SM throat the wavefunction $Y_2(\lambda^S\tdz) + Q_m^S
J_2(\lambda^S\tdz)$ drops off roughly as as $\exp(-b\tdz)$. This can be seen from the asymptotic form
of the Bessel functions, namely 
\beqa
  \sqrt{\pi x/2} J_\nu(x) & \approx & \cos(x-\nu \pi/2 - \pi/4) \, , \labell{AsymptoticJ} \\
  \sqrt{\pi x/2} Y_\nu(x) & \approx & \sin(x-\nu \pi/2 - \pi/4) \, . \labell{AsymptoticY}
\eeqa
Indeed, taking into account~\reef{QmSWithSink} we find
\beqa
  \psi(z) &\approx& \frac{2N_m^S}{\pi\cos(\lambda^S\tdz_S)} \cos[\lambda^S(\tdz-\tdz_S)] \nn \\
          & \sim & N_m^S e^{-b\tdz} \, .  \labell{AsymptoticPsi}
\eeqa
Note that $N_m^S$ remains finite in the limit $z_S \ra \infty$.

This is an interesting result: it says that as long as the
wavenumber $\lambda_S^A$ has an imaginary part (which will be the case whenever $S^S \neq S^A$), the
A throat KK modes will effectively penetrate the SM throat only to distance at most $z_p \sim 1/b$, no
matter how long the throat is. 

An important consequence of the finite penetration length, if it is shorter than the throat length $z_S$, is that it
can lead to a substantial reduction of the overlap integrals~\reef{zetan} that are the coupling constants in the decay
rates. We will now argue that the decay rate~\reef{GammaA2S} is so fast that it implies a penetration length $z_p$ much
smaller than the length of the SM throat; hence, in a sense, the rate $\Gamma_{A,2S}$ is inconsistent and we must find
a way of estimating the physically realized decay rate.

To obtain the penetration length $z_p$ implied by the decay rate $\Gamma_{A,2S}$, we need to find the 
sink $S^S$ that corresponds to $\Gamma_{A,2S}$. This problem is solved in Appendix~\ref{app:SinksFromDecayRates}; the
result is
\beq
  b \sim \frac{S^S}{m} \sim k \frac{k^3}{M_5^3} \sigma_A^3 \sigma_S^{-2} \, , \labell{naiveb}
\eeq
so 
\beq
  \frac{z_S}{z_p} \sim \frac{k^3}{M_5^3} \sigma_A^3 \sigma_S^{-3} \, . \labell{naivezp}
\eeq
Clearly, unless $k/M_5$ is extremely small, we find 
\beq
  z_p \ll z_S \, ,
\eeq
meaning, as we claimed above, that the penetration length is much shorter than the throat length and hence the
physically realizable decay rate $\hat \Gamma_{A,2S}$ will be smaller than the result~\reef{GammaA2S}. We now estimate
$\hat \Gamma_{A,2S}$.

\subsection{Estimating the effective decay rate}

We have an A throat mode that decays by interactions that are localized in the S throat. We argued above that its
wavefunction will decay exponentially as a function of (coordinate) distance in the S-throat. The situation can be
described as a reservoir of probability density in the A throat that slowly leaks into the S throat, where the
probability density decays. Clearly, the probability decay rate cannot be faster than the rate of leakage; in fact,
since the probability does not accumulate in the throat either, the decay and leakage rates must equal.

The leakage rate is given by the probability current 
\beq
  j_z \sim \frac{i}{m} \psi^* \stackrel{\leftrightarrow}{\partial} \psi \,  \labell{Current}
\eeq
evaluated at the top of the throat. The current~\reef{Current} will consist of two terms $j=j_1+j_2$, with $j_1$ proportional to the
imaginary part of the coefficient $Q_m^S$, while $j_2$ is proportional to the imaginary part of the wavenumber $\lambda^S$. Let us first look
at $j_1$: using the asymptotic forms of the Bessel functions and  $b\tdz_S>1$, we find
\beq
  Q_m^S \approx -i
\eeq
so the wavefunction in the SM throat is actually the outgoing Hankel function $H_2^+(\lambda \tdz)$! The parameter
$\lambda \tdz$ is complex, so the amplitude of the wavefunction decays exponentially, as we have shown in the previous
section; near the top of the throat we find
\beq
  j_1 \sim k\sigma_A^5 \sim m \sigma_A^4 \, ,
\eeq
namely the rate at which plane waves tunnel between two infinite throats, considered in~\cite{dimopoulos0106, barnaby,
gary, kofman}. The part $j_2$ of the current can be
estimated as
\beq
  j_2 \sim - j_1 \left| \frac{\Im \{ \lambda^S \}}{\Re {\lambda^S}} \right| \, .
\eeq  
The negative sign is always present and means that the decay of the wavefunction further slows down the
``leaking'' of the probability density into the SM throat. An important question now is whether $j_2$ is comparable to
$j_1$, that is whether the imaginary and real parts of the SM throat wavenumber $\lambda^S$ are comparable. If 
it were so, the decay rate could be significantly suppressed even compared to the plane wave tunnelling
rate (this possibility was also pointed out in~\cite{kofman}).
As we argued earlier, the imaginary part of the wavenumber determines the penetration length $z_p$ as $z_p^{-1} \sim
\Im \{ \lambda^S \}$. In the following subsection we will estimate the latter and show that, at least in our
approximation, the current $j_2$ is indeed negligible, so the effective decay rate roughly equals the plane wave tunnelling
rate,
\beq
  \hat \Gamma_{A, 2S} \approx \Gamma_{tun} \sim m \sigma_A^4 \, .  \labell{SelfconsistentRate}
\eeq
Thus, while the naive decay rate~\reef{GammaA2S} is not the physical one because it is larger than the tunnelling rate, 
it does have a physical meaning: it is in turn small enough that current $j_2$ remains negligible compared to $j_1$.

\subsection{Estimating the penetration length}

We would like to calculate the penetration length $z_p \equiv b^{-1}$ of the A throat KK mode wavefunction into the SM
throat self-consistently, at least within our approximations. The strategy is to calculate the decay rate
$\hat \Gamma_{A,2S}(z_p)$ as a function of
$z_p$, and equate it to the maximum physical rate~\reef{SelfconsistentRate}.

The calculation will follow the same line as in Section~\ref{sec:Decay}, with the following modifications. First, we
will assume that the decaying wavefunction, instead of decreasing exponentially with the characteristic length $z_p$,
equals the naive wavefunction for $z<z_p$, and is zero for $z>z_p$; this means the overlap integral~\reef{zetan} will
only extend to $z=z_p$. The wavefunctions entering the integral are still approximated by~\reef{ApproxIKKPsiSMIR}
and~\reef{ApproxSMKKPsiSMIR} and the result is (cf.~\reef{CplS_A2S})
\beq
  \hat \zeta^S_{A,2S} \sim \hat \xi \frac{k^{1/2}}{M_5^{3/2}} w_0 \frac{m_0^2}{k^2}  \sigma_A^{1/2}
     \sigma_S \sigma(z_p)^{-5/2} \, . \labell{HatCplS_A2S1}
\eeq
The correction factor $\hat \xi$ is now, in analogy with the discussion below~\reef{CplS_A2S}, roughly
\beq
  \hat \xi \sim \sigma_A^{-1} \sigma(z_p) \, , \labell{hatxi}
\eeq
so the improved coupling can be written as
\beq
  \hat \zeta^S_{A,2S} \sim \frac{k^{1/2}}{M_5^{3/2}} w_0 \sigma_A^{3/2} \sigma_S \sigma(z_p)^{-3/2} \, .
       \labell{TdCplS_A2S}
\eeq
Calculating the decay rate also involves summing over accessible final states. Since the wavefunction of the decaying mode
only penetrates the SM throat to distance $z_p$, it will only couple (appreciably) to SM throat modes whose mass is
large enough for them to not be suppressed for $z<z_p$, so we will only include modes of mass $m>1/z_p$ in the sum over
final states. Roughly speaking this means that the number of accessible final states changes from $z_S m_0$ to $z_p
m_0$. Applying the results of Appendix~\ref{app:DecayRates} we find
\beq
  \hat \Gamma_{A, 2S} (z_p) \sim \frac{k^4}{M_5^3} \sigma_A^8 \sigma_S^2 \sigma^{-5}(z_p) \, .
    \labell{hatGammaA2S}
\eeq
Equating $\hat \Gamma_{A, 2S} (z_p)$ with the actual rate~\reef{SelfconsistentRate} then gives
\beqa
  \frac{\sigma(z_p)}{\sigma_A} &\sim& \left[ \frac{k^3}{M_5^3} \frac{\sigma_S^2}{\sigma_A^2} \right]^{1/5} 
          \labell{sigmaz_p} \\
  & \ll & 1 \, .
\eeqa
Put in words, we find that the penetration length $z_p$ is longer than the A throat length $z_A$. Hence the
imaginary part of the wavenumber in the SM throat, $\Im \{ \lambda^S \} \sim z_p^{-1}$, is much smaller than the real
part $\Re \{\lambda^S\} \sim \tdz_A^{-1}$, which is what we wanted to show.

\mysection{Level repulsion and crossing during SM throat relaxation}

\label{sec:LevelCrossing}

Recall that according to the conjecture of~\cite{rob}, the SM throat will be shortened during inflation by the
effects of interactions of the A and the SM throat length modulus. After inflation ends, the SM throat
should relax to its ``full'' length. While the quantitative details of the relaxation, such as its rate, are
unknown, one can imagine modelling it via a time-dependent SM throat length $z_S$ in the doubled RS model that we
employ to represent the full string theory setup.

A changing SM throat length leads to an
additional possible mechanism of energy transfer. As
the length of the SM throat increases, it will periodically go through an interval where the lengths of the
A and SM throats are tuned in the sense that there is a (near) degeneracy between pairs of one level from each
throat. As is usual in quantum mechanics, the energy levels of coupled systems avoid crossing each other, which in this
case means that a level that started out as an A throat KK mode will, after passing through the tuning point,
become an SM throat KK mode and vice versa. This mechanism could, if the SM throat relaxation is slow enough, ``suck'' the
energy out of the A throat and into the SM throat, where the KK modes would quickly decay into lower-lying states
that cannot switch back to the A throat. We start by presenting a simple version of this
mechanism that neglects the presence of complex parts in the frequency and wavenumbers induced
by mode decay; then we investigate whether properly accounting for the said complex parts
changes the picture qualitatively.  

We will assume that the SM throat relaxation is adiabatic, at least as far as the A throat KK modes (and
SM throat modes of comparable masses) are
concerned. The usual condition for adiabaticity, $ \dot \omega / \omega^2 \ll 1 $ implies, via $\tdz_S
\omega \approx \const$, a condition $\dot z \ll \sigma_A/\sigma_S$. We will simply assume this condition is
satisfied. Let us now look at how the KK modes evolve with a changing $z_S$ while $z_A = \const$. To get oriented, we
first take the sink-less case of real $\lambda^A = \lambda^S = m$. The spectrum-generating matching
condition~\reef{QmS} can be re-written in a more symmetric form as
\beq
  \frac{Y_1 + Q_m^A J_1}{Y_2 + Q_m^A J_2} + \frac{Y_1 + Q_m^S J_1}{Y_2 + Q_m^S  J_2} = 0 \, ,
           \labell{MatchingCondnSinkFree}
\eeq
where the Bessel functions are all evaluated at $m/k$ and the coefficients $Q$ are given by
\beq
  Q_m^{A,S} = - \frac{Y_1(m\tdz_{A,S})}{J_1(m\tdz_{A,S})} \, .
\eeq
Let us start with $z_S$ such that the two throats are detuned (that is, only one of the coefficients $Q$ is large).
Recall that an A throat KK mode has a large $Q_m^A$ and a small ($\sim 1$) $Q_m^S$. As $\tdz_S$ increases,
the mode's mass remains approximately constant until $Q_m^S$ starts becoming large as well. At this point the mass
$m$ starts moving such that $m\tdz_S$ will remain approximately constant, $Q_m^A$ will become small ($\sim 1$) and
$Q_m^S$ will become large -- what started as an A throat KK mode has become an SM throat mode. The
switch-over will repeat itself in reverse when $z_S$ decreases further and the throats become tuned again. Of course,
if we had started by tracking an SM throat mode, it would switch over into the A throat at the first point of
tuning, and then back to the SM throat at the second one. This is an example of a standard quantum-mechanical level
repulsion: at each point where the two throats have tuned lengths, the levels of one A throat mode and one
SM throat mode avoid crossing by switching their ``home throat''. In principle, this switching can facilitate energy
transfer between the throats; if it is slow enough to allow the modes to fully relax, a KK mode particle originally located
in the A throat will relocate to the SM throat where it can decay to lower KK modes very efficiently. If the
A throat mode is the low-lying one, it is quite likely that the products of its decay will have masses below
the lowest lying A throat KK mode and hence will not be subject to switching anymore.

%One is that the mode switching cannot happen faster than it takes light
%to travel from the A throat into the SM one; this time is of order $t_{sw} \sim z_S = 1/(k\sigma_S)$. To
%beat the decay into gravitons, the switch-over must be faster than the decay time into gravitons $t_g =
%1/\Gamma_{A,2g}$ with the rate given by~\cite{gary}
%\beq
  %\Gamma_{A,2g} \sim k g_s^{-1/4} \left( \frac{k}{M_5} \right)^{15/8} \sigma_A^3 \, . \labell{GammaA2g}
%\eeq
%Requiring $t_sw \ll t_g$ 

This simple picture has an obvious loophole: when one takes into account the presence of the sink(s), the mode wavenumbers
$\lambda^{A,S}$ acquire (different) imaginary parts and it becomes possible for the levels to avoid each other in
the complex plane even though the real parts cross. It is easy to see that this will happen when the imaginary part
of one of the wavenumbers is large: in that case the corresponding coefficient $Q_m$ cannot become
large (as we show below),
%\footnote{The asymptotic form of $Q_m$ is $Q_m = -Y_1(\lambda \tdz)/J_1(\lambda \tdz) \approx \cot(\lambda \tdz
%+ \const)$; if the imaginary part of $\lambda$ is nonzero, $\cot (\lambda \tdz)$ is bounded from above by $\sim
%[1 + \exp(\Im \{\lambda \tdz \})]/\exp(\Im \{\lambda \tdz \})$.}
meaning that the mode can never switch to that side.  On the other hand, one would 
expect that if the sink(s), and hence the imaginary parts of the wavenumbers, are very small, the system should
behave similarly to the one without sinks at all (in other words, the equations governing the eigenvalues are
analytic and we do not expect any discontinuities for $S \ra 0$). Hence there should be a maximum sink that still
allows the mode switching to take place; above that sink the real parts of the wavenumbers can cross and modes remain
in their respective throats. Let us find the condition for the switching to take place, and see whether our model
satisfies it.

The spectrum is determined by~\reef{RSJumpWithSink}. It is clear that as long as both the (real) mass $m$ and the
sink $S$ are much smaller than the curvature scale $k$, the only way to satisfy~\reef{RSJumpWithSink} is to have at
least one of the coefficients $Q_m$ large. Suppose we start with detuned throats and pick an A throat mode,
\ie $Q_m^A$ is large: to leading order,
\beq
  Q_m^A \approx - \frac{\lambda^S Y_1^SY_2^A + \lambda^AY_1^A Y_2^S}{\lambda^AJ_1^AY_2^S} \sim \frac{k^2}{\lambda^2}
    \, ,  \labell{QmIApprox}
\eeq
where we have introduced a shortened notation for the Bessel functions by denoting $Y_1^S \equiv Y_1(\lambda^S/k)$
etc, and in the last expression we have neglected the difference between the wavenumbers. As we have noted
before,~\reef{QmIWithSink} implies that $\lambda^A$ must be (nearly) real.

For the mode to switch throats, $Q_m^S$ must become of the same order at the tuning point (and then remain large
afterwards). The expression~\reef{QmSWithSink} reduces, in the asymptotic regime, to
\beq
  Q_m^S \approx - \cot(\lambda^S \tdz_S + \const) \, , \labell{QmSApprox}
\eeq
where the $\const = -\pi/4$ can be neglected for the purposes of our argument. Let us denote $\lambda^S\tdz_S = a_S
+ ib_S$ with $a_S, b_S$ real. We then have
\beq
  Q_m^S \approx - \frac{\sin a_S\cos a_S -i \sinh b_S \cosh b_S}{\sin^2 a_S + \sinh^2 b_S} \, . \labell{QmSApprox2}
\eeq
It is clear that if $b_S$ is of order 1 or larger, $Q_m^S$ will always be of order 1. Hence $b_S$ will have to be
small; we can then approximate $\sinh b_S \approx b_S$, $\cosh b_S \approx 1$. For small $b_S$ the
expression~\reef{QmSApprox2} will be largest when $\sin a_S \approx \b_S$, and the maximum value attained will be
$\sim 1/ b_S$ (both the real and imaginary parts will attain values of this order). Comparing with the required order
of magnitude~\reef{QmIApprox} we find that switching can only occur when $b_S \lesssim |\lambda^2|/k^2$. 
This means that the imaginary part of the wavenumber itself must be tiny:
\beqa
  \Im \{ \lambda^S \} & = & \frac{b_S}{\tdz_S} \sim \frac{|\lambda_S^2|}{k^2 z_S} \nn \\
              & \lesssim & k \sigma_A^2 \sigma_S  \sim \frac{\sigma_A^2}{\tdz_S} \, ,  \labell{MaxImLambda}
\eeqa
where we have used that the typical value of the wavenumber is $\sim k\sigma_A$. This condition is not satisfied by our
model: while~\reef{MaxImLambda} implies that the penetration length $z_p \sim (\Im \{ \lambda^S \})^{-1}$ must be much
longer than the SM throat length $z_S$, at the end of Section~\ref{sec:ModesWithSink} we have argued that the
penetration length will be much shorter than the throat length (and the ``self-consistent'' penetration
length obtained from~\reef{sigmaz_p} does indeed come out shorter than the SM throat length). Hence we conclude that
the mode switching cannot occur.

\mysection{Enhancing KK mode tunnelling between throats}

\label{sec:Tunnelling}

In previous sections we have found that, generically, energy transfer between the throats will proceed at the rate
given by the plane wave tunnelling rate of~\cite{dimopoulos0106}. 
In this section we look for ways of enhancing the tunnelling rate by modifying the
effective potential barrier in the UV region of the RS model. In standard
RS the tunnelling probability of a mode of mass $m$ (with $mL \ll 1$) is suppressed as~\cite{dimopoulos0106}
\beq
  P_{RS} \sim (mL)^4\, .  \labell{P_RS}
\eeq
Throughout this section we will, for convenience, use curvature radii $L$ instead of curvature scales $k$;
they are straightforwardly related by
\beq
  L \equiv \frac{1}{k} \, .
\eeq
Since we are interested in plane wave tunnelling rates, we will work with the plane wave boundary
conditions that are commonly used in tunnelling
calculations, \ie the SM side wave function will be an asymptotically 
purely outgoing wave $\sqrt{m\tdz} H_2^+(m\td z)$. On the A side we will have a mix
of incoming and outgoing waves (with a nonzero net incoming flux), $\sqrt{m\tdz} H_2^-(m\td z)$
and $\sqrt{m\tdz} H_2^+(m\tdz)$ respectively, where 
\beq
  H_\nu^{\pm}(x) = J_\nu(x) \pm i Y_\nu(x)
\eeq
are the usual Hankel functions. Unlike the $\ZZ_2$ boundary conditions employed in previous sections,
the tunnelling ones do not explicitly depend on the location of the end-branes. We will not take into account the
effects of decay here; the decay effects can be obtained by a straightforward adaptation of the results of previous
sections.

Our aim of lowering the potential barrier will be achieved by replacing the (single) UV brane
with tension~\reef{BraneTension} by a certain number, $N$, of branes with smaller tension
\beq
 T' = T/N\, . \labell{SmallerTension}
\eeq
Such ``generalized Randall--Sundrum'' setups were considered, \eg in~\cite{oda9908, hatanaka9909}. 
If $N$ is small, say 2,
tunnelling can be speeded up by resonant effects akin to the resonant tunnelling phenomenon
known from ordinary quantum mechanics (resonant tunnelling was also discussed by~\cite{henry,
henry0602}); we give an explicit example of such a setup in
Section~\ref{sec:ResonantTunnelling}. On the other hand, a large number of appropriately placed
branes will lower the effective potential
barrier and hence enhance tunnelling, as we show in Section~\ref{sec:EquidistantBranes}. 
All cases will have in common the fact that the
curvatures on the two sides of any particular membrane differ; therefore we will start with
a general discussion of the matching conditions for such situations.

\subsection{Many-brane Randall--Sundrum}

\label{ManyBraneRS}

Let us assume that our system consists of $N$ AdS regions bounded by $N+1$ branes
located at $z=z_i, i=0,..,N$. The brane at $z_0$ will be taken to be the annihilating
brane that are responsible for inflation, while the one at $z_{N+1}$ will be the SM brane.
In each region between the branes the metric is the standard AdS
metric~\reef{background}, but the warp factor will have different coefficients in each
region, namely
\beq
  \sigma_i(z) = \frac{L_i}{z+\td L_i} \, . \labell{ithWarpFactor}
\eeq
The variables $L_i$ are (up to sign) the curvature radii of each region. The sign of $L_i$
will be positive (negative) if the warp factor decreases (increases) with increasing $z$.
The shifts $\td L_i$ are chosen such that the warp factor is continuous at every membrane.
The actual values of $\td L_i$ depend on a choice of the origin of the $z$ coordinate and
the normalization of the warp factor. 

The curvature jump conditions at each of the branes require that the reduced 
tension $\td T_i \equiv T_i/M_5^3$ of brane $i$ and the curvature parameters $L_i$ and
$L_{i+1}$ on both sides of the membrane satisfy (cf.~\reef{BraneTension})
\beq
  \td T_i = 12\left( \frac{1}{L_{i+1}} - \frac{1}{L_i} \right) \, . \labell{ithJumpCondn}
\eeq
This condition is valid both at the interior branes as well as at the inflating and SM
branes, the only difference being that at the latter two we also impose $\ZZ_2$
boundary conditions on the background.

The mode equation in each region has the same form as in the single UV brane case,
namely~\reef{PsiEom}, but with the potential $V(z)$ now given by 
\beq
  V(z) = \frac{15}{8}\frac{1}{\left(z+\td L_i\right)^2} \, , \ \ \ \ z_{i-1} < z < z_i\, .
           \labell{ithPotential}
\eeq
Each brane will contribute a term of the form
\beq
  V_i^{(b)} = \td T_i \sigma(z_i)\delta(z-z_i)
\eeq
into the effective potential as well.

Away from the branes in each region the KK wave functions $\psi_m(z)$ are given by a
general combination of the Bessel functions $J_2(m\td z_i(z))$ and $Y_2(m\td z_i(z))$,
where
\beq
  \td z_i(z) \equiv |z+ \td L_i|  \, . \labell{ithtdz}
\eeq
At each membrane the wave function $\psi_m$ must be continuous,
\beq
  \psi_m(z\ra z_i^-) = \psi_m(z\ra z_i^+)\, , \labell{ithPsiCondn}
\eeq
and its derivative must obey the jump condition
\beq
 -\frac{1}{2\psi_m(z_i)} \left[ \prt_z \psi_m(z_i^+) - \prt_z \psi_m(z_i^-) \right] =
  \frac{\td T_i}{16} \sigma(z_i)\, .\labell{ithPsiJumpCondn}
\eeq
The zero mode wave function is a direct generalization of~\reef{ZeroModePsi}: in $i$-th
region it is
\beq
  \psi_0(z) = N_i\frac{1}{(\td z_i(z))^{3/2}} \, , \labell{ithZeroModePsi}
\eeq
where the normalization constants $N_i$ are chosen such that the wave function is continuous
across every membrane.\footnote{The proper jump in the first derivative then follows
automatically.} In fact, since the warp factor $\sigma$ has the same functional form and
coefficients that do make it continuous, we can also write 
\beq
  \psi_0(z) = N_0 \sigma^{3/2}(z) \, ,
\eeq
where the normalization constant $N_0$ is now the same for all regions.

\subsection{Throats with different curvature radii}

\label{sec:DifferentCurvatures}

An interesting generalization of the models of~\cite{barnaby, dimopoulos0106} is to allow the curvature
radii of the two throats to differ. The radii can in principle arbitrary as long as the tension of the brane
that separates them obeys the jump condition~\reef{ithJumpCondn}. Let us first look at the plane wave
tunnelling rate through a one-brane potential barrier that separates two regions with curvature radii $L^A$
and $L^S$, respectively. Normalizing the warp factor to 1 at the UV brane ($z=0$) leads to $\td L^A= L^A, \ \td L^S =
L^S$ (see~\reef{ithWarpFactor}). The KK mode wavefunctions on the A and SM sides are,
respectively,
\beqa
  \psi^A(z) &=& \sqrt{m\tdz_A} \left[ AH_2^-(m\tdz_A) + B H_2^+(m\tdz_A) \right] \, , \labell{DiffCurvPsiI}\\
  \psi^S(z) &=& \sqrt{m\tdz_S} C H_2^+(m\tdz_S) \, , \labell{DiffCurvPsiS}
\eeqa
where 
\beq
  \tdz_{A,S} = |z + L^{A,S}| \nn \, .
\eeq
As usual, on the A side we have a superposition of an incoming and a reflected wave, whereas on the
SM side only the transmitted component is present. Continuity and first derivative jump
at the brane,~\reef{ithPsiCondn} and~\reef{ithJumpCondn}, lead to
\beq
  \frac{C}{A}  = \sqrt{\frac{L^A}{L^S}} \frac{H_2^-(mL^A)H_1^+(mL^A) - H_1^-(mL^A)H_2^+(mL^A)}
                                           {H_2^+(mL^S) H_1^+(mL^A) + H_2^+(mL^A) H_1^+(mL^S)}\, .
               \labell{DiffCurvTunnelAmpl}
\eeq
Let us first assume that the curvature radii are such that $mL^A \ll 1, mL^S \ll 1$. Then, using the
small-argument expansions of Bessel functions~\reef{SmallArgument}, we find
\beq
  \left| \frac{C}{A}\right| \sim m^2 (L^A)^{3/2}(L^S)^{1/2} \, . \labell{DiffCurvSmallRadii}
\eeq
Hence, making $L^S$ moderately larger  than $L^A$ can increase the plane wave tunnelling rate somewhat, but
not by a lot.

Next, let us assume that the SM curvature radius is large, such that $mL^S \gsim 1$. The Hankel functions can
then be characterized as $ H_\nu^\pm(mL^S) \sim e^{\pm i mL^S}/\sqrt{mL^S}$, and we find  
\beq
  \left| \frac{C}{A}\right| \sim (mL^A)^{3/2} \, . \labell{DiffCurvSmallLargeRadii}
\eeq
In particular, this formula also applies when $L^S\ra \infty$, that is when the SM side of the UV membrane
is flat, and so characterizes the tunnelling of KK modes out of a single AdS$_5$ throat. Compared to the
equal radii case $L^A = L^S$, the tunnelling rate $P = |C/A|^2$ is clearly enhanced by a factor of
$1/(mL^A)$.

Having significantly different radii in
the two throats has other interesting consequences: for example, it can relax the
constraint~\reef{SigmaSConstraint} on the relative sizes of the 
warp factors at the A and SM branes -- the derivation of the constraint from~\reef{zSConstraint}
(which remains valid even in the case of different curvature radii) assumed equal curvature radii on both
sides. Conversely, if the curvature radius $L^S$ of the SM throat were much larger than $L^A$, the SM warp
factor $\sigma_S \equiv L^S/(L^S+ z_S)$ could be much larger than 
the bound~\reef{SigmaSConstraint} while allowing KK modes to tunnel from the A throat to the SM one.
Phenomenologically, however, such a setup may not be desirable because of the presence of a large hierarchy
between the curvature scales in two throats; it may also be difficult to realize such a setup as a string
theory compactification.

\subsection{Resonant tunnelling through a gravity box}

\label{sec:ResonantTunnelling}

One possibility to speed up the energy transfer between throats is to use an analog of
the resonant tunnelling effect from standard quantum mechanics: split the single potential
barrier (the UV region) in the RS geometry into two, with a flat region (a well with zero
potential in the effective Schr\"odinger equation) inbetween~\cite{lykken}. If the frequency of the
tunnelling wave is tuned to the size of the of the well in the middle, the tunnelling
rate can approach one. As argued in~\cite{henry, henry0602} who analyzed a similar setup, in a realistic
string theory compactification one would expect that the two throats are joined by a bulk region that is
modelled by the flat interval.

\begin{figure}[ht]
\begin{center}
\begin{tabular}{cc}
\epsfig{file=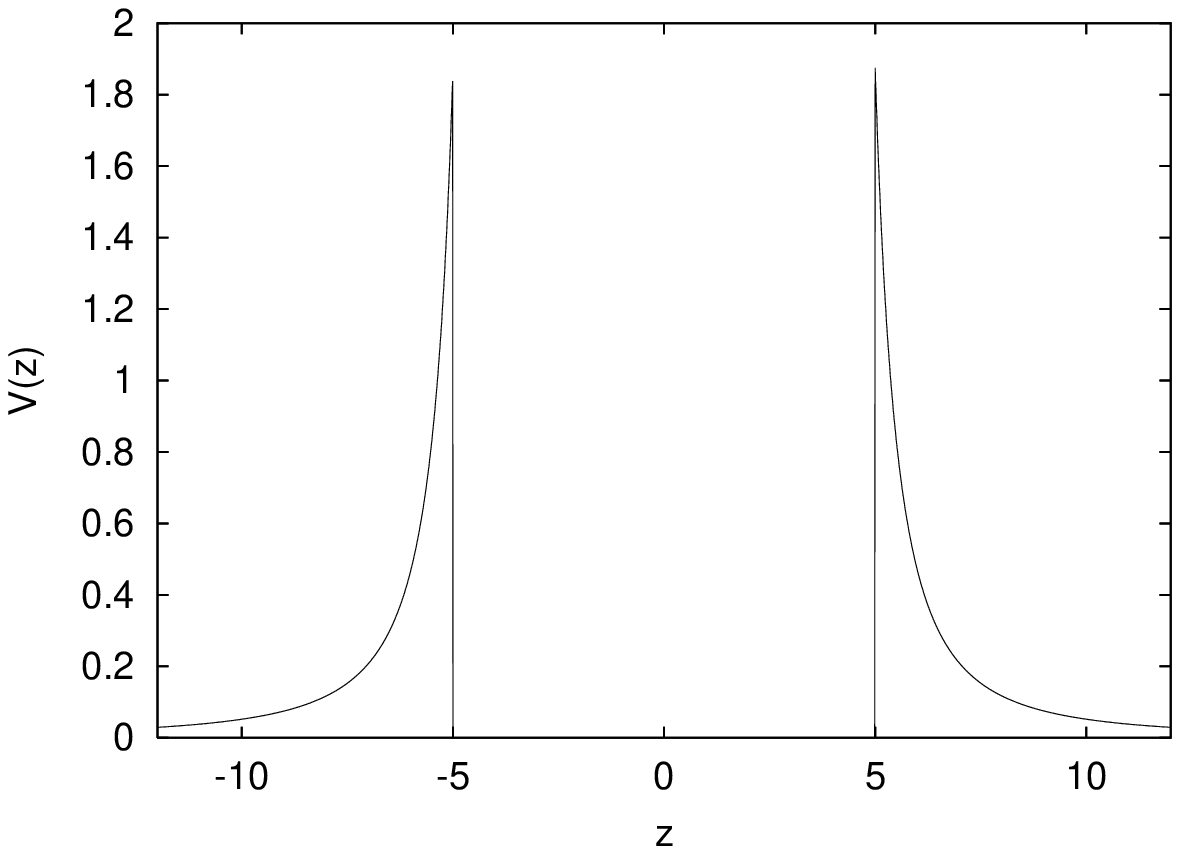 ,width=7.5cm} & \epsfig{file=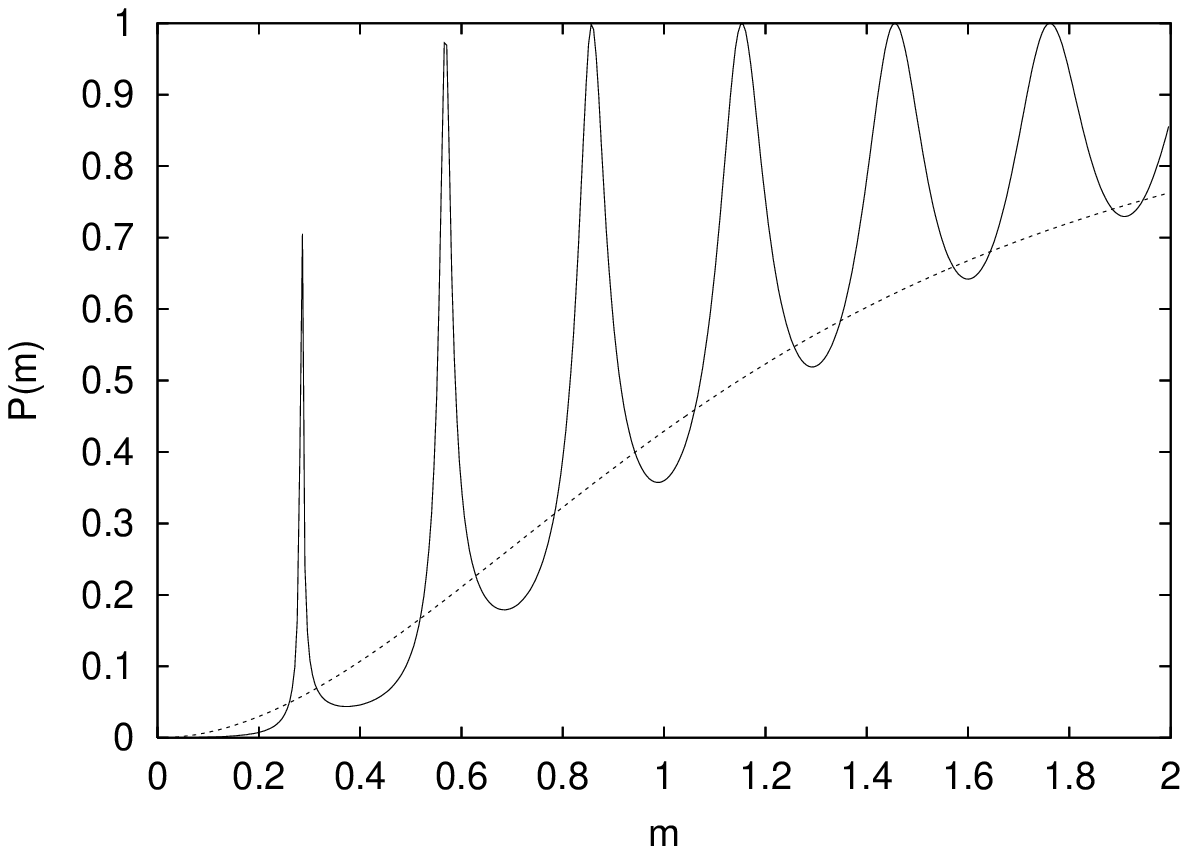,width=7.5cm} \\
(a) & (b)
\end{tabular}
\end{center}
\caption{Two branes with a flat region between them. Graph (a) plots the potential $V(z)$, while
(b) shows the tunnelling rate $P$ as a function of the KK mode mass $m$ (solid line) together with
the reference tunnelling rate of the standard double RS geometry (dotted line). The brane
separation is $2l=10$. Tunnelling is greatly enhanced for masses that fall within
the resonant tunnelling peaks. Note also the suppression of the tunnelling rate for very light modes.}
\label{fig:box}
\end{figure}

Hence we consider a setup of the type described in~\cite{lykken}, but with the tunnelling
boundary conditions. The geometry
consists of two AdS throats, whose curvatures we will take to be the same for simplicity,
separated by a flat region. The curvature radius of the throats will be denoted by $L$,
while the length of the box in the middle will be $2l$, with the branes sitting at $z=\pm
l$. The effective potential is plotted in Fig~\ref{fig:box}~(a). The warp factor is constant in the central
region and will be normalized to 1 there (so the coordinate and
proper length of the middle box coincide). On the A side (region 1) the wave
function of a KK mode with mass $m$ is
\beq
  \psi_m^{(1)}(z) = \sqrt{m\td z_1(z)} \left[ AH_2^+(m\td z_1(z)) + BH_2^-(m\td z_1(z))
                                       \right], \labell{BoxPsi1}
\eeq
while in the middle box (region 2) the wave function is
\beq
  \psi_m^{(2)}(z) = C\cos (mz) + D\sin (mz)\, , \labell{BoxPsi2}
\eeq
and on the SM side (region 3) we have
\beq
  \psi_m^{(3)}(z) = \sqrt{m\td z_3(z)} E H_2^+(m\td z_3(z)) \, . \labell{BoxPsi3}
\eeq
In the notation of Section~\ref{ManyBraneRS} the parameters $L_i$ are 
\beq
  L_3 = -L_1 = L, \ \ \ L2=\td L_2 \ra \infty\, ,
\eeq
while the shifts are (from requiring $\sigma(z=\pm l) = 1$)
\beq
  \td L_1 = -\td L_3 = -L +l \, .
\eeq
It is straightforward, if somewhat tedious, to solve the matching
conditions~\reef{ithPsiCondn} and~\reef{ithPsiJumpCondn}. To shorten the result, let us
denote 
\beq
  t \equiv \tan ml \, ,
\eeq
and omit the arguments of all Hankel functions, since they are all evaluated at $mL$. The
tunnelling amplitude $E/B$ is then
\beq
  \frac{E}{B} = \half \frac{ \left( t+\frac{1}{t}\right) \left( H_2^-H_1^+ - H_1^- H_2^+
                               \right)} 
                           { {H_2^+}^2 - {H_1^+}^2 + \left(\frac{1}{t}-t \right)H_2^+
                             H_1^+ }\, . \labell{ResonantTunnelAmplitude}
\eeq
As a check, when $t\ra 0$, so $1/t$ dominates over everything else, the above tunnelling amplitude
reduces to the bulk-less expression given in~\cite{dimopoulos0106}.

The tunnelling resonance occurs when the denominator becomes small, \ie when
\beq
  \frac{1}{t} - t \approx \frac{ {H_2^+}^2 - {H_1^+}^2 }{H_2^+ H_1^+ }\, .
             \labell{ResonantCondn}
\eeq
If we concentrate on KK modes whose mass $m$ is small compared to the inverse curvature
radius, $mL \ll 1$, we can use the asymptotic forms~\reef{SmallArgument} of the Bessel
functions, and we find that the denominator becomes small when 
\beq
  \frac{1}{t} - t \sim mL \ll 1\, ,
\eeq
that is when $t \approx 1$. This is not surprising -- it means that to get an
amplification of the tunnelling rate, the size of the central
box must be roughly the same as the wavelength of the tunnelling particle. As a numerical
illustration, we plot the potential and the tunnelling rate of this setup in
Fig.~\ref{fig:box}. The parameters used in the plot are $l=5$, $L=1$. 

This is bad news for phenomenology. According to the discussion around
Eq.~\reef{EffectiveBraneScale} we would expect the masses of the produced KK modes to be
smaller than the annihilation scale $M_A$, which in turn is expected to be a few orders of
magnitude lower than the inverse curvature radius $L$ of the throats. Hence, the flat
region would have to be unnaturally long as compared to the characteristic length  of the
throats, and presumably as compared to the radius of the internal manifold in a string
theory compactification; in other words, the hierarchy problem that the warping was
supposed solve would reappear in a different guise.

Lastly, one can consider the regime of $t$ small but not too small; more precisely, 
\beq
  1 \gg t \gg \frac{Y_1(mL)}{Y_2(mL)} \sim mL \, . \labell{BulkRegime}
\eeq
In this regime the tunnelling amplitude behaves roughly as
\beq
  \frac{E}{B} \sim \frac{mL}{t} (mL)^2 \, , \labell{BulkRegimeTunnelAmpl}
\eeq
meaning that the tunnelling rate with bulk present is suppressed by an additional factor of $(mL/t)^2$ compared
to the bulk-less tunnelling rate. The extra suppression is also evident in the plot~\ref{fig:box}(b).

\subsection{Comments on the WKB approximation}

Incidentally, our results~\reef{DiffCurvSmallLargeRadii} and~\reef{BulkRegimeTunnelAmpl} provide a cautionary
note on the WKB methods used, \eg by~\cite{henry, henry0602}. In the WKB approximation, when the tunnelling
amplitudes are small, the tunnelling amplitude $\Theta_{tt}$ between two throats is roughly given by the square of the
rate $\Theta_{tb}$ at which particles can tunnel from one throat into (flat) bulk,
\beq
  \Theta_{tt} \approx \Theta_{tb}^2 \, . \labell{WKBSquare}
\eeq
It is clear that such formulae must be used with caution: for example, the tunnelling amplitude between two
RS throats, calculated in~\cite{dimopoulos0106}, is $\Theta_{tt} \sim (mL)^2$, while the throat-to-bulk
tunnelling amplitude~\reef{DiffCurvSmallLargeRadii}
is $\Theta_{tb} \sim (mL)^{3/2}$. Further, when the two
throats are separated by a flat bulk region, the tunnelling amplitude~\reef{BulkRegime} does formally exhibit 
the $(mL)^3$ behaviour that one would expect from $\Theta_{tb}$,
but it also depends on the length of the bulk region represented by the parameter $t$ (and the dependence is
very different from the WKB approximation as used in~\cite{henry, henry0602}).

\subsection{Lowering the potential barrier by multiple branes}

\label{sec:EquidistantBranes}

Another way of speeding up the tunnelling process is to lower the potential barrier in the
UV region by replacing the single UV brane by a large number of branes of lower tension
that are also spatially separated along the $z$ direction. The basic idea is to utilize
the fact that the (squared) warp factor $\sigma_i^2(z)$ and the potential $V(z)$, given
respectively in~\reef{ithWarpFactor} and~\reef{ithPotential}, differ by a factor of the
(squared) curvature radius $L^2$. By arranging the branes such that the the warp factor
becomes of order 1 only when the curvature radius is also large, we can achieve a
significant lowering of the effective potential barrier between the throats.

We have not attempted an exhaustive search for the configuration most favourable for
tunnelling. It is clear nevertheless that the the modification must happen in the region where the warp
factor (and the Randall--Sundrum potential) are large, namely around $z=0$. In this section we
present a toy example of such a modification.

Suppose we arrange $N=2n$ branes with tension $\td T_1= \td T/N$ such that the middle
region is (approximately) flat. Then the curvature radius of a region $i$ satisfies
\beq
  \frac{1}{L_i} = 12 \td T_1 (i-n) \, .
                       \labell{ithEquidistL}
\eeq
As discussed above, $1/L_i^2$ is precisely the factor (up to $15/8$) by which the potential and
the warp factor differ:
\beqa
  V_i(z) &=& \frac{15}{8} \frac{\sigma_i^2(z)}{L_i^2} \nn \\
         &=& \frac{15}{8} \left( 12 \td T \sigma_i(z) \right)^2 \frac{(i-n)^2}{N^2} \, .
          \labell{ithEquidistPotential}
\eeqa
The last expression makes the suppression of the potential more apparent: the warp factor
is largest in the middle (flat) region, but the potential there is suppressed by the
$(i-n)^2/N^2$ factor. Because the shifts $\td L_i$ entering the warp factor are determined
recursively, it is not possible to derive a simple formula for the maximum of the
potential~\reef{ithEquidistPotential}, so we give a few simple examples for illustration.

In our example, we fix the warp factor value at the maximum to be 1. This choice  ensures
that the warping at the ``top'' of the throats is the same for all setups --- while arbitrary,
this has the physical consequence of measuring masses at the top of the throats in the same
units. Further, since we consider a many-brane scenario to be a near-the-top modification of
the ``standard'' model of~\cite{barnaby, dimopoulos0106}, we fix the curvature and warp factors at the bottom
of the throats to be the same in all cases. 

\begin{figure}[ht]
\begin{center}
\begin{tabular}{cc}
\epsfig{file=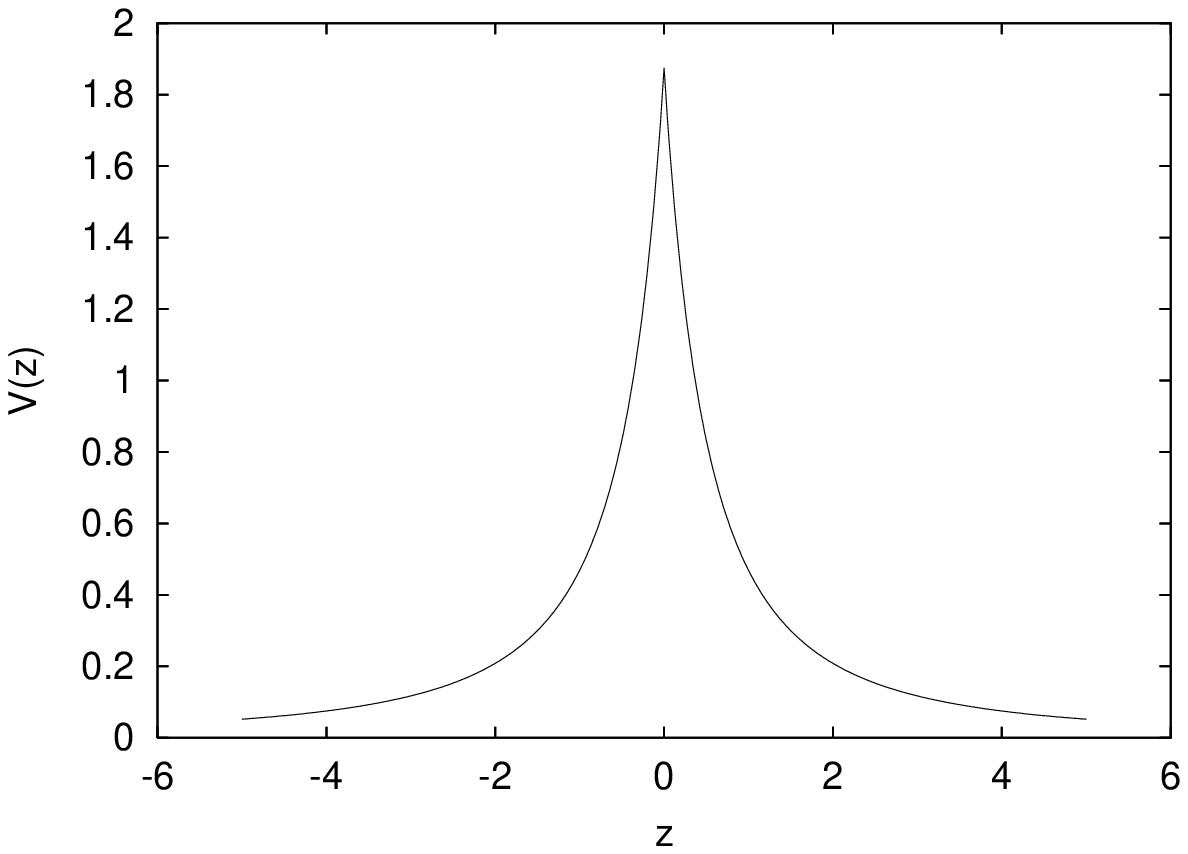 ,width=8cm} & \epsfig{file=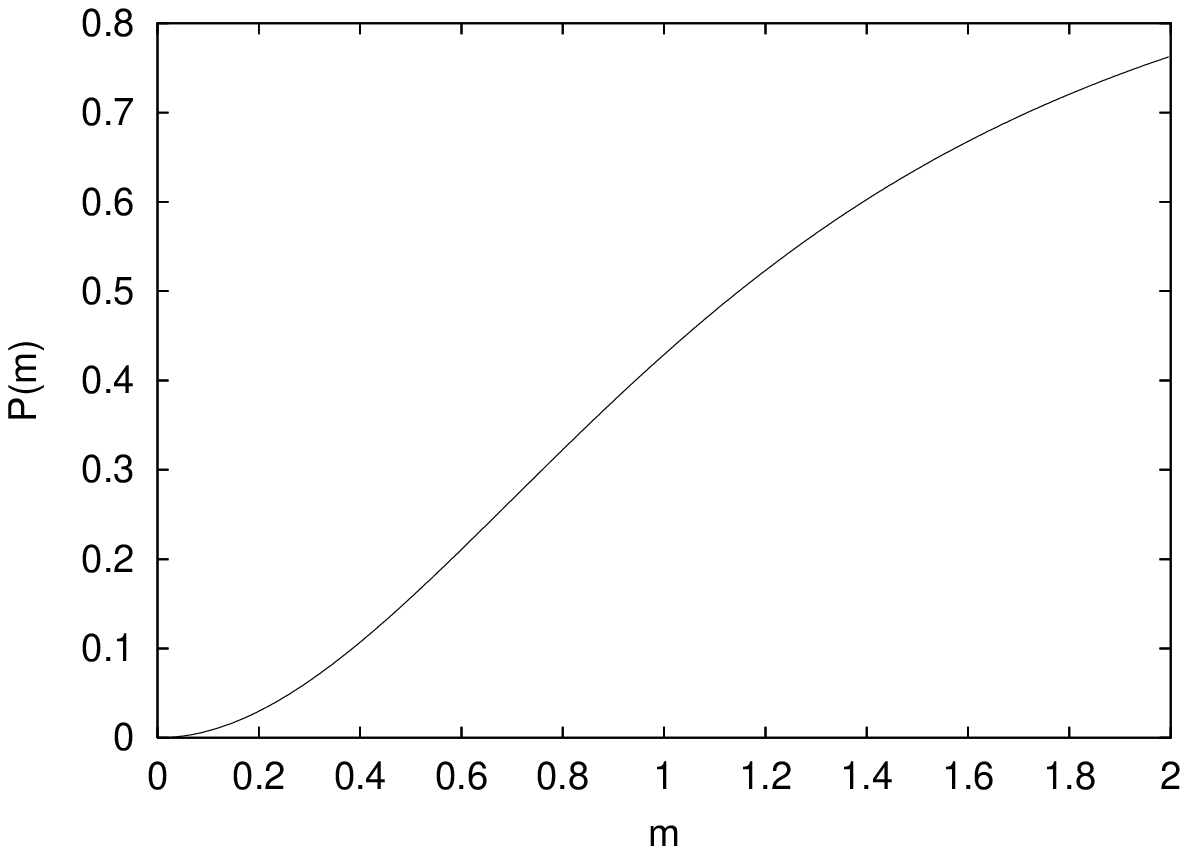,width=8cm} \\
(a) & (b)
\end{tabular}
\end{center}
\caption{The single central brane setup with constant curvature. Graph (a) plots the potential $V(z)$, while
(b) shows the tunnelling rate as a function of the KK mode mass $m$. }
\label{fig:single}
\end{figure}

In Fig.~\ref{fig:single} we show potential as a function of the
coordinate $z$ and the tunnelling rate as a function of the mass $m$ of the tunnelling KK mode for the
benchmark doubled RS (\ie constant-curvature, single central brane) setup of~\cite{barnaby,
dimopoulos0106}. As expected, the tunnelling rate becomes substantial when the KK mode mass
$m^2$ becomes the same as the potential at the top of the barrier.

\begin{figure}[ht]
\begin{center}
\begin{tabular}{cc}
\epsfig{file=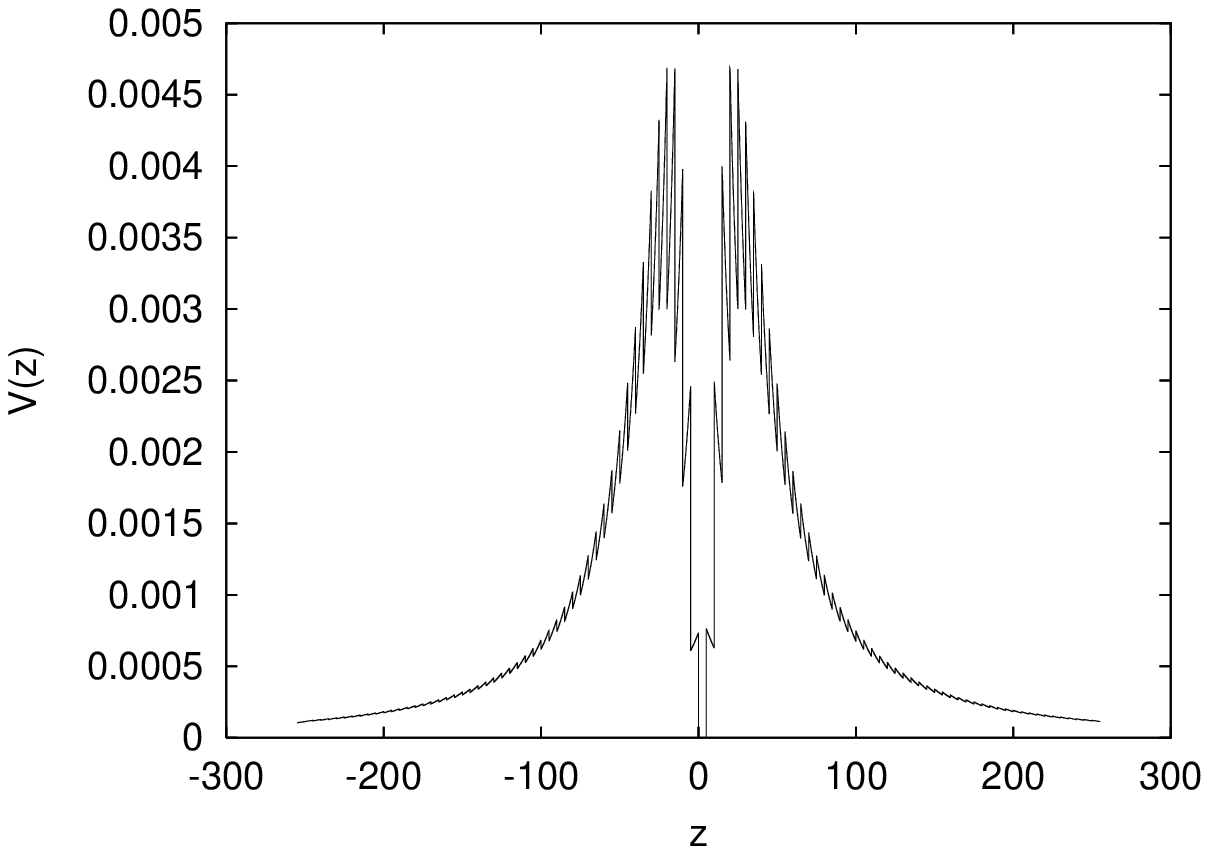 ,width=8cm} & \epsfig{file=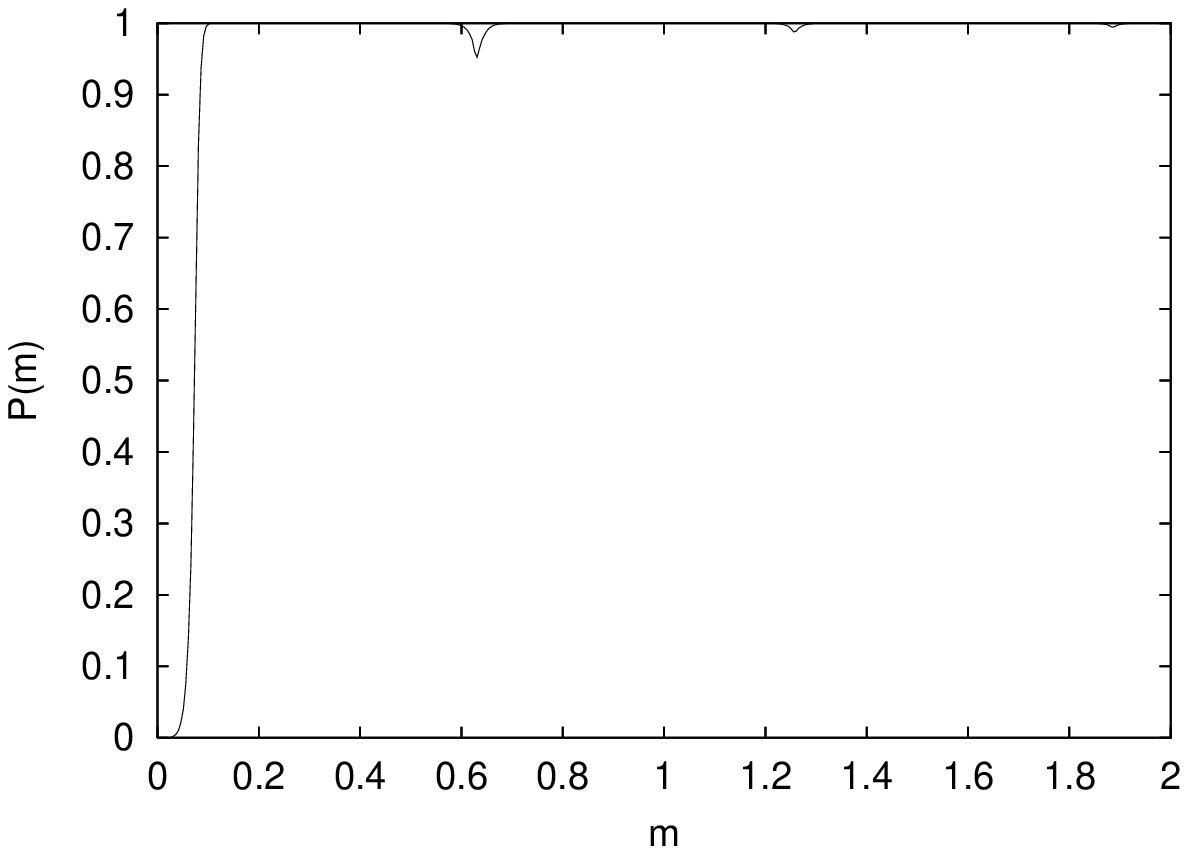,width=8cm} \\
(a) & (b)
\end{tabular}
\end{center}
\caption{The multiple brane setup. Graph (a) plots the potential $V(z)$, while
(b) shows the tunnelling rate as a function of the KK mode mass $m$. }
\label{fig:multi}
\end{figure}

\begin{figure}[ht]
\begin{center}
\begin{tabular}{cc}
\epsfig{file=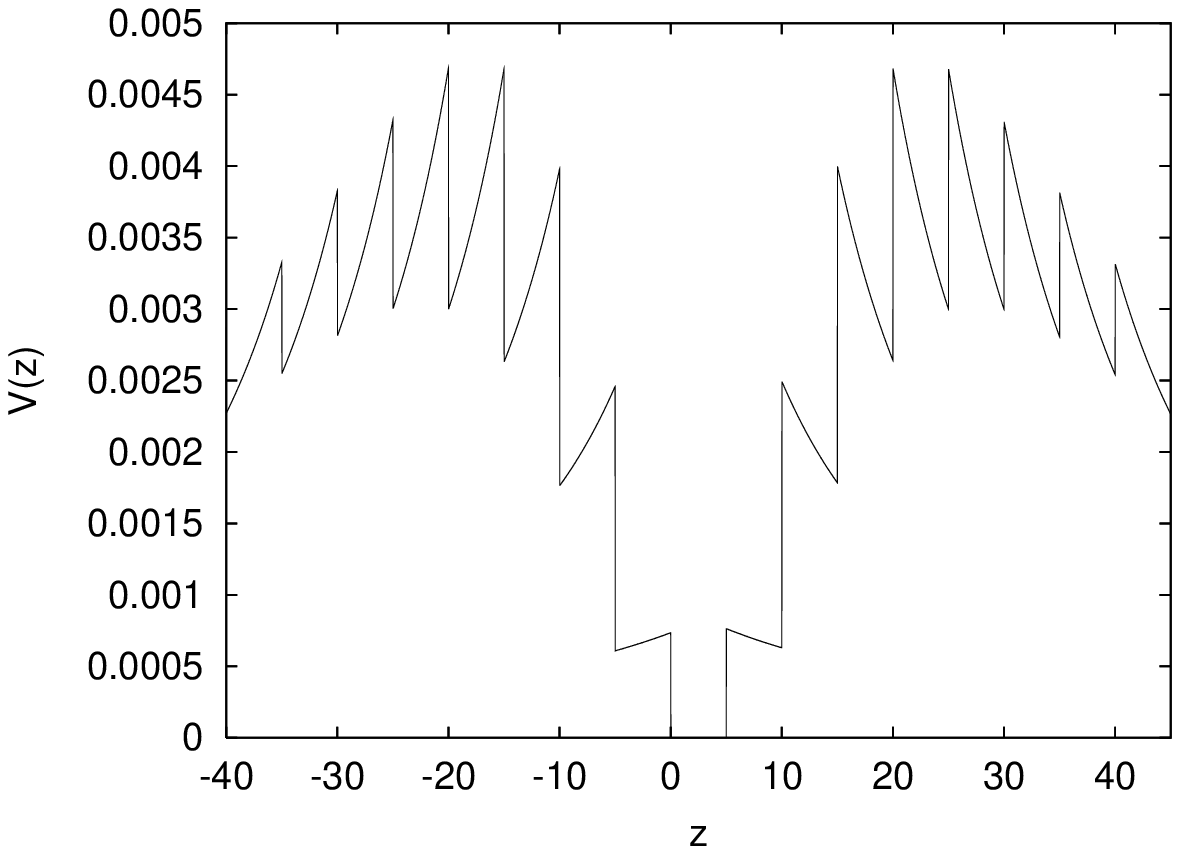 ,width=8cm} & \epsfig{file=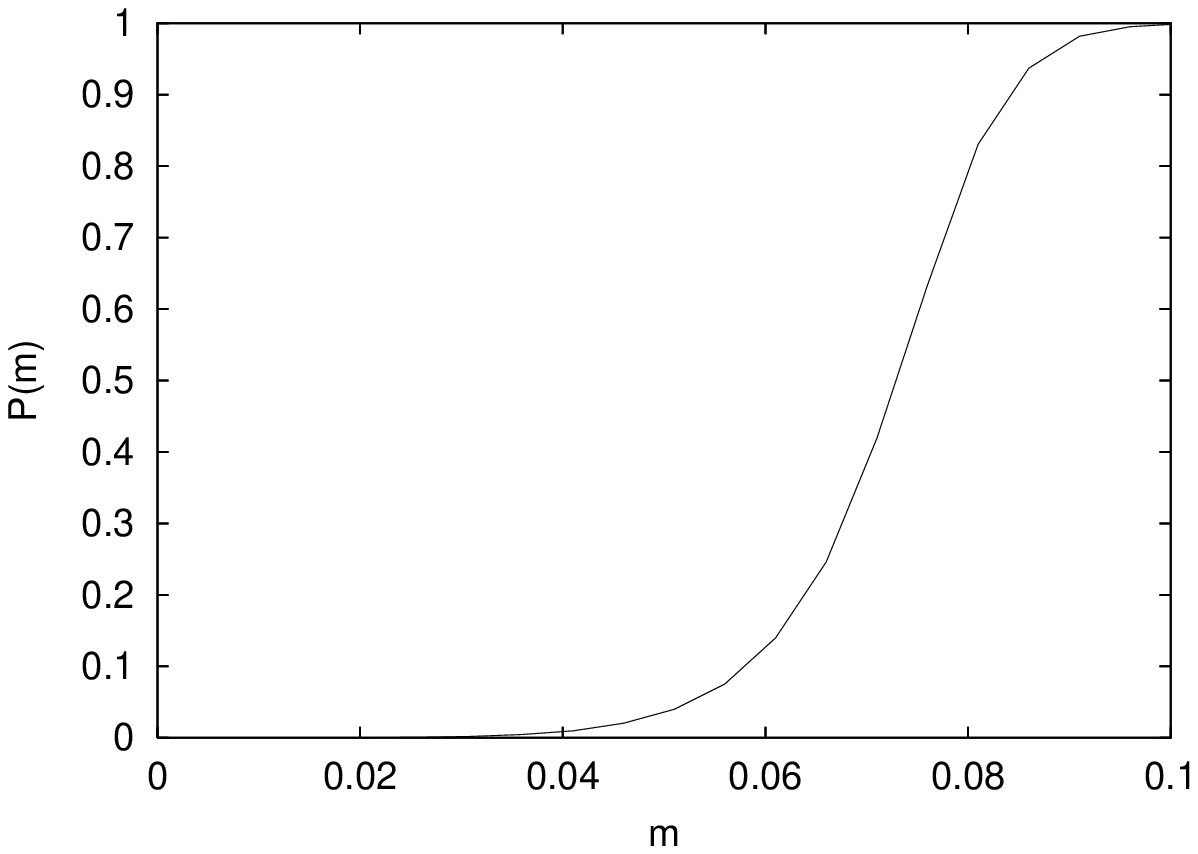
,width=8cm}  \\
(a) & (b)
\end{tabular}
\end{center}
\caption{Detail of (a) the potential $V(z)$ and (b) the tunnelling rate $P(m)$ shown in Fig.~\ref{fig:multi}.}
\label{fig:multi2}
\end{figure}

Now let us look at a setup where the central brane is replaced by 100 equidistant branes (50 on each side of
$z=0$) with tension equal 1/100 of the tension of the single brane. The potential and the tunnelling rate are
shown in Fig.~\ref{fig:multi}. In Fig.~\ref{fig:multi2} we zoom in on the central region to show the
potential in more detail. The potential barrier is significantly lower and the tunnelling rate for low-lying
KK modes is correspondingly enhanced.

Are the implications of this toy model applicable to an honest string theory compactification?
Recent work~\cite{henry, henry0602} studying string theory throats seem to answer this
question in the negative. We would like to take a more cautious position: it seems clear that the
tunnelling rates depend to a large degree on the details of the gluing of the throat to the
bulk; these details that are not yet under sufficient control.
In fact, the effective potentials for the KK modes in~\cite{henry, henry0602} and
our potential plotted in Fig.~\ref{fig:multi} are qualitatively similar, but their quantitative
details are apparently sufficiently different to cause a major difference in tunnelling
rates. 

\mysection{Conclusions}

\label{sec:Conclusions}

We have presented an exhaustive discussion of the properties of the KK modes that are responsible for
the energy transfer from the A throat to the SM throat. We have found that tunnelling in the
conventional sense will proceed only if the SM throat is substantially longer than the A throat,
namely only for $z_S \gsim z_A (k\tdz_A)^4$. 
%
%In any case, the usual tunnelling rate is too slow in the
%sense that most KK modes would decay into gravitons before tunnelling into the SM throat, unless the
%parameters of the model (or the string theory compactification) take special values~\cite{gary, henry0602}.
%
Even when the A throat KK modes do not tunnel in the conventional sense, their wavefunction has
a tail in the SM throat that is big enough to induce a large decay rate into 2 SM throat KK modes.
Naively this rate should increase with the length of the SM throat and, for a long SM
throat, will be large enough to make most of the annihilation KK mode energy go into SM throat KK modes.
However, a more careful analysis taking into account the effects of wavefunction decay shows that the
decay rate into two SM throat KK modes cannot get larger than a certain maximum. The effect can be
understood intuitively from the fact that a large decay rate would shorten the penetration length of the
A throat KK mode into the SM throat, which in turns leads to a suppression of the overlap of its
wavefunction with the SM throat KK modes, and that in turn suppresses the decay rate. We have shown that the physically
realized decay rate is effectively given by the plane wave tunnelling rate. Nevertheless, the naive decay rate is {\em
not} completely meaningless: we show that it is turn small enough that it does not induce a further slowing down of the
tunnelling/decay rates via the complex frequency effect discussed in~\cite{kofman}.

An additional mechanism of energy transfer we have investigated is based on level repulsion and can be
pictured as throat switching by KK modes when the spectrum of the SM throat is gradually lowered by the
relaxation of the SM throat. We found that the presence of complex parts in the mode frequencies destroys
the level repulsion effect, because levels can now avoid each other in the complex plane; hence, the energy
transfer by switching cannot occur. 
 
%In the course of evaluating various ways of energy transfer we have repeatedly found that the energy
%transfer could be speeded up relative to decay into gravitons if the ratio $k/M_5$ were small. In this
%context $k$ has the meaning of the UV barrier height; however, the reason why $k/M_5$ cannot be made too
%small is that $k$ is also related to the compactification scale. Clearly, most of the negative
%conclusions of Sections~\ref{sec:Modes}--\ref{sec:LevelCrossing} would be avoided if one could lower the
%barrier height while keeping the 
%compactification scale unchanged; hence, in the next section, we look for ways to achieve this. 

We have then proposed a simple modification of the UV region of the geometry that, at least in the
phenomenological 5-dimensional model, leads to a drastic enhancement of tunnelling rates between the two
throats. The modification consists of replacing a single UV brane with an array if branes with lower tension
such that the curvatures deep in the throats remain the same. This has the effect of lowering the potential
in the effective Schr\"odinger equation that the KK modes have to tunnel through, leading to an enhancement
of their tunnelling rates. It is not clear, especially in the light of the recent studies of
stringy throats~\cite{henry, henry0602}, whether a similar mechanism could be realized in an
honest string compactification; we argued that a better knowledge of the geometry where the
throats are glued to the bulk is necessary to answer this question definitely. 

\section*{Acknowledgements}

It is a pleasure to thank Rob Myers for many key discussions and suggestions, and Xingang Chen, Jean Francois
Dufaux, Hassan Firouzjahi, Lev Kofman, Dmitry Podolsky, Gary
Shiu, Henry Tye and Bret Underwood for interesting discussions. The author is supported by an NSERC Discovery grant. Research
at the Perimeter Institute is supported in part by funds from NSERC of Canada and MEDT of Ontario. 

\appendix

\mysection{Modes of a two-throat system}

\label{app:Suppression}

In this appendix we study in detail the modes of the two-throat system described in
Section~\ref{sec:Suppression}. 

\subsection{Small mass modes: $mz_A \ll 1$}

\label{sec:SmallMassModes}

Taking the mass to be small allows us to use standard formulae for Bessel functions
of small arguments, namely
\beqa
  Y_1(x) \approx  -\frac{2}{\pi} \frac{1}{x} & \ \ \ \ & Y_2(x) \approx  -\frac{4}{\pi} \frac{1}{x^2} \nn \\
  J_1(x)  \approx  \frac{1}{2} x & \ \ \ \ & 
  J_2(x)  \approx  \frac{1}{8} x^2 \, . \labell{SmallArgument}
\eeqa
We then have
\beqa
  Q_m^A & \approx & \frac{4}{\pi} \frac{1}{(m \td z_A)^2} \labell{ApproxQmI} \\
  Q_m^S & \approx & -\frac{8}{\pi} \frac{k^2}{m^2} \labell{ApproxQmS} \\
  N_m^S & \approx & \frac{\pi}{8} \left( \frac{m}{k} \right)^2 \labell{ApproxNmS} \\
  N_m^A & \approx & N_m^S \labell{ApproxNmI} \\
  \psi_m^A(z_A) & \approx & \frac{1}{2} \sqrt{\frac{m}{k}} \frac{1}{(k\td z_A)^{3/2}}
                             \labell{ApproxPsimI}
\eeqa
and find
\beq
  R_\psi \equiv \frac{\psi_m^A(z_A)}{\psi_0(z_A)} \approx \frac{1}{2} \frac{\sqrt{m}}{k} \, .
                  \labell{ApproxPsiRatio}
\eeq
One should not be surprised that this ratio is not dimensionless; after all, we are
comparing wavefunctions that have different (namely continuum vs. discrete)
normalizations. Physically one cannot talk about a single mass value (or mode) out of a continuum;
one must always integrate over a certain range of masses to get a meaningful result. One
can, \eg integrate the probability density of finding a KK mode of mass $m'<m$ at the
A brane  and compare it to the probability density of finding the zero mode
there, \ie calculate
\beq
  p(m) = \frac{ \int_0^m dm' \, \left(\psi_{m'}^A(z_A)\right)^2 }{\psi_0^2(z_A)} \, .
                 \labell{DefnOfP}
\eeq
Using~\reef{ApproxPsiRatio} we find
\beq
  p(m) \approx \frac{1}{4} \frac{m^2}{k^2} \, . \labell{PForSmallM}
\eeq
As derived, this expression is only valid for $m \ll 1/z_A$. We would like to extend
this calculation to modes whose mass is of order $1/z_A$, because we expect that the
effective brane scale $M_A$ will be roughly of that magnitude.

\subsection{Medium mass modes: $mz_A \sim 1$}

\label{sec:MediumMassModes}

There is no good approximation formula for the Bessel functions when their arguments are
of order 1; the best one can do is to say that generically, all of the relevant Bessel
functions are also of order 1. Thus one cannot say much more about $Q_m^A$ than that it
will also generically be of order 1. That is enough, however, to fix $Q_m^S$ by the jump
condition at the Planck brane (recall that we still assume $mk \ll 1$, so all Bessel
functions evaluated at the Planck brane can be replaced by their small-argument
approximations). We find that $Q_m^S$, and therefore also the normalization constants
$N_m^S$ and $N_m^A$ are given by the same expressions as in the small mass case,
namely~\reef{ApproxQmS}---\reef{ApproxNmI}. The wavefunction of the massive mode at the
A brane is then\footnote{
We use the $\sim$ sign to denote our ignorance of factors of order 1 (we reserve $\approx$ for approximations
that hold up to small corrections).}
\beq
  \psi_m^A(z_A) \sim \frac{m^{5/2} \td z_A^{1/2}}{k^2} \labell{ApproxPsimI2}
\eeq
Interestingly enough, the ratio $\psi_m^A(z_A) / \psi_0(z_A)$ is then (using $m\td
z_A \sim 1$) roughly the same as in the small mass case, namely
\beq
  R_\psi = \frac{\psi_m^A(z_A)}{\psi_0(z_A)} \sim \frac{\sqrt{m}}{k} \, .
                       \labell{ApproxPsiRatio2}
\eeq
We show below that this result holds only for truly ``generic'' values of $m$; we will
see below that at
special points, the ratio $R_\psi$ can grow to be much larger than the above value.

The result~\reef{ApproxPsiRatio2} implies that one can integrate up
to $m\sim z_A$ in~\reef{DefnOfP} with result
\beq
  p(m) \sim \left(\frac{m}{k}\right)^2 \, , \labell{PForMediumMass}
\eeq
that is the non-zero mode amplitude at the A brane is suppressed by roughly the
square of the ratio of the inflation and string scales (modulo the radius of the 5th
dimension). 

\begin{figure}[ht]
\begin{center}
\begin{tabular}{cc}
\epsfig{file=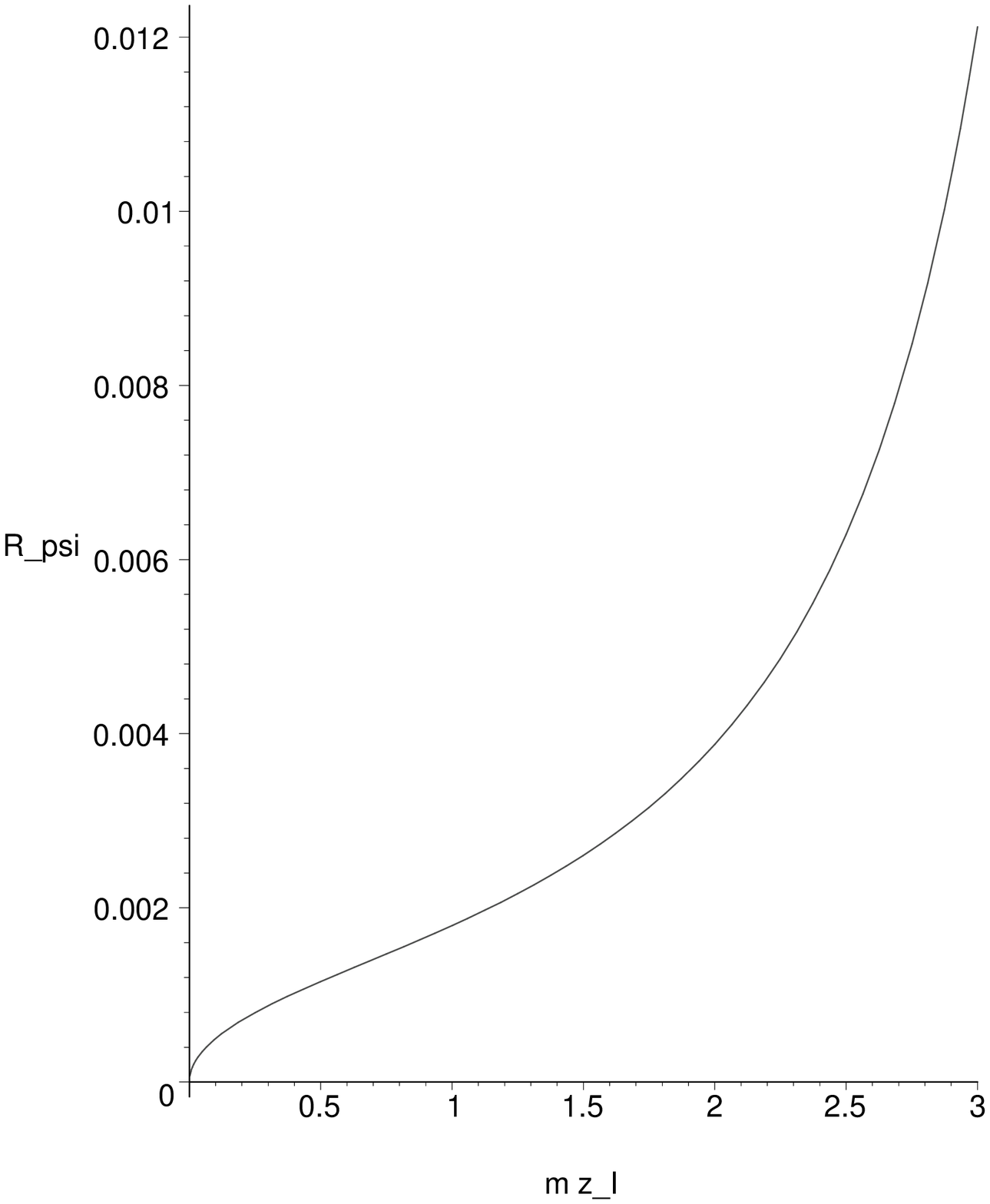 ,width=8cm} & \epsfig{file=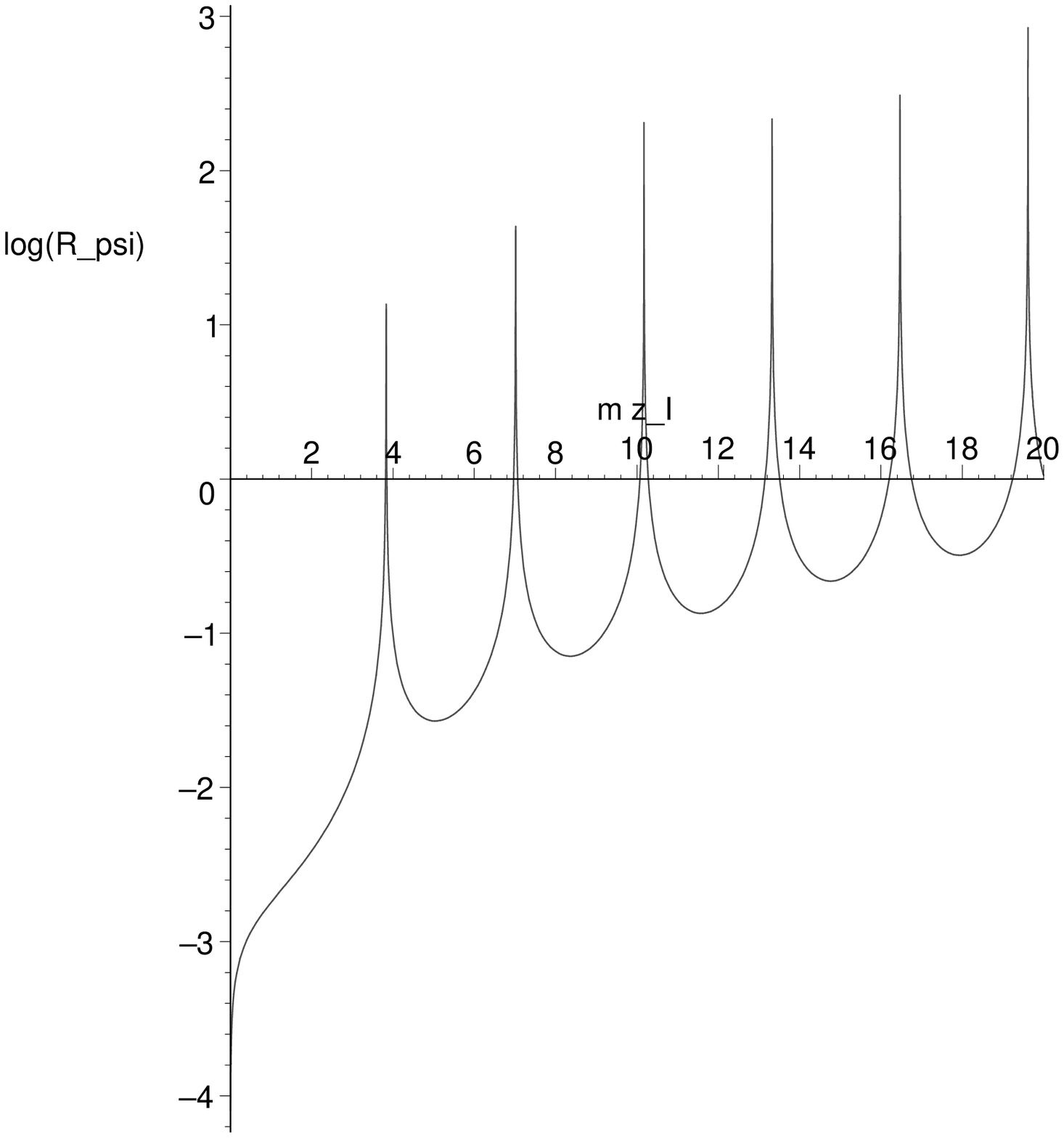
,width=8cm}  \\
(a) & (b)
\end{tabular}
\end{center}
\caption{Plot (a) shows the ratio $R_\psi = \psi_m^A(z_A)/\psi_0(z_A)$, as a function of $m$ (in units
of $1/z_A$). For small
$m$ the ratio behaves as $ \sim \sqrt{m}/k$. In this plot $k=1, z_A=10^5$. In (b) we plot the logarithm of the ratio
$R_\psi = \psi_m^A(z_A)/\psi_0(z_A)$, as a function of $m$ (in units
of $1/z_A$), for a larger range of $m$. This plot is logarithmic, so the KK mode wavefunctions are suppressed
with respect to the
zero mode whenever the curve is below the horizontal axis. Note the presence of spikes that actually turn
out to dominate the integrated ratio $p(m)$.}
\label{fig:RatioOfPsi2}
\end{figure}

Since in this case we do not have factors of order one under control, it is
useful to plot the $\psi_m^A(z_A)/\psi_0(z_A)$ numerically. The plot can be found in
Fig.~\ref{fig:RatioOfPsi2}(a).

%\begin{figure}[ht]
%\begin{center}
%\epsfig{file=RatioOfPsi2.eps ,width=10cm}
%\caption{The ratio $R_\psi = \psi_m^A(z_A)/\psi_0(z_A)$, as a function of $m$ (in units
%of $1/z_A$). For small
%$m$ the ratio behaves as $ \sim \sqrt{m}/k$. In this plot $k=1, z_A=10^5$.
%}
%\label{fig:RatioOfPsi2}
%\end{center}
%\end{figure}
%
%\begin{figure}[ht]
%\begin{center}
%\epsfig{file=RatioOfPsi2-0-20.eps ,width=10cm}
%\caption{Logarithm of the ratio $R_\psi = \psi_m^A(z_A)/\psi_0(z_A)$, as a function of $m$ (in units
%of $1/z_A$), for a larger range of $m$.
%This plot is logarithmic, so the KK mode wavefunctions are suppressed with respect to the
%zero mode whenever the curve is below the horizontal axis. Note the presence of spikes that actually turn
%out to dominate the integrated ratio $p(m)$.}
%\label{fig:RatioOfPsi2-0-20}
%\end{center}
%\end{figure}

To see that the validity of our approximations is restricted to ``generic points'', let
us look at plots of the ratio $R_\psi$ for values of $m$ that are several
times $1/z_A$. The plot in Fig.~\ref{fig:RatioOfPsi2}(b)
shows that for certain
masses, the wave function of the corresponding modes on the A brane is parametrically 
amplified. One might suspect that these spikes in $R_\psi$ appear when either $Q_m^A$ 
diverges (at the zeros of
$J_1(m\td z_A)$), or the denominator in the expression~\reef{NmI} for $N_m^A$ becomes
zero, and $R_\psi$ would diverge. In fact, it turns out that neither of these special points 
can cause the spike; rather, the spikes appear at values for which $Q_m^S$ approaches
zero, and $R_\psi$ remains bounded (but large). For $m\ll k$ one can use the approximate 
values~\reef{SmallArgument} of the Bessel functions to
show that at the points where $Q_m^S =0$, the KK mode wave functions attain values $\sim
k^2/m^2$, and the ratio $R_\psi$ grows to $R_\psi \sim (k/m)^{7/2}/\sqrt{k}$. 

Physically, the interesting question is for what values of $m$ does the integrated
probability density ratio $p(m)$ of~\reef{DefnOfP} reach values $~\sim 1$. The presence of the spikes makes
a numerical integration difficult, so we provide estimates that will turn out to be good enough. We start
with generic values of $m$ and consider the contribution of the spikes separately below. 

For a values of the KK mode mass $m$ away from the spikes, the approximate form~\reef{ApproxPsimI2} holds as
long as $m < k$. Keeping $z_A$ in the expressions explicitly, we arrive at
\beq
  R_\psi \sim \frac{\sqrt m}{k} \left( mz_A \right)^2,
\eeq
and 
\beq
  p(m) \sim \frac{m^2}{k^2} \left(mz_A\right)^4 \, . \labell{ApproxPm}
\eeq
This result is counterintuitive: it says that for the KK modes to dominate over the zero mode, their mass
must be so large that they are already far into their asymptotic regime, $m \gg 1/z_A$. However, in the
asymptotic regime we would expect the KK modes to dominate over the zero mode, because the KK modes approach
plane waves with constant magnitude, whereas the zero mode falls off as a power of $z$. We seem to be
missing some KK mode contributions; this discrepancy is a hint that the spikes contribute significantly, and
actually dominate $p(m)$. Let us now show that.

As we have already mentioned, the spikes appear at masses $\bar m$ such that $Q_{\bar m}^S = 0$. We remind
the reader that we still assume $\bar m\ll k$. From~\reef{QmS}
we find that the corresponding $Q_{\bar m}^A$ must be
\beq
  Q_{\bar m}^A \approx -2\frac{Y_1(\bar m/k)}{J_1(\bar m/k)} \sim \frac{k^2}{\bar m^2} \gg 1\, .   \labell{QmISpike}
\eeq
This equation together with~\reef{QmI} determines the values of $\bar m$ where the spikes occur; the first
few spikes will be at $\bar m \sim 1/z_A \ll k$ in accord with our assumption. It is also worth noting that
the spike occurs when $J_1(m\td z_A)$ is small but not zero, $J_1(m\td z_A) \sim 1/Q_{\bar m}^A \sim
\frac{\bar m^2}{k^2}$.  From~\reef{NmS} we find
\beq
  N_{\bar m}^S = 1 \, ,
\eeq
and from~\reef{NmI}, taking into account~\reef{QmISpike}
\beq
  N_{\bar m}^A \approx N_{\bar m}^S \, .
\eeq
To estimate the contribution to $p(m)$ from a spike, we will find its width $\delta m$ at half maximum. The
relevant quantity here is $R_\psi^2(m) = \psi_m^2(z_A)/\psi_0^2(z_A)$, so half maximum is the point where the
KK mode wavefunction squared $\psi_m^2$ drops to half its peak size. The drop can come from two sources
(see~\reef{GeneralPsi}): either $N_m^A$ or $Q_m^A$ can halve. Let us for now assume that $N_m^A$ varies
faster, find the width $\delta_m$ and then show that the variation in $Q_m^A$ due to $\delta_m$ is smaller.

Hence, we are looking for $\delta_m$ such that for $m=\bar m + \delta m$, $(N_m^A)^2 = (N_{\bar m}^A)^2/2$.
This implies $(N_m^S)^2 = (N_{\bar m}^S)^2/2$ and hence
\beq
  Q_m^S =1 \, .
\eeq
Keeping only the leading terms in~\reef{QmS} (recall again $m/k \ll 1$) we find that $Q_m^A$ must change
from the $-2Y_1(m/k)/J_1(m/k)$ of~\reef{QmISpike} to
\beq
  Q_m^A \approx \frac{2Y_1(m/k) - J_1(m/k)}{J_1(m/k)} \approx Q_{\bar m}^A - 1 \,  \labell{QmIHalfMax}
\eeq
that is $\delta Q_m^A \approx -1$. Recall that $Q_{\bar m}^A$ is large because $J_1(\bar m\td z_A)$ is near its
zero; thus $\delta Q_m^A$ will come from the change in $J_1(m\td z_A)$ caused by $\delta m$, while $Y_1(m\td
z_A)$ remains approximately constant. Denoting the change in $J_1(m\td z_A)$ by $\delta J_1$, we require
\beq
  \frac{Y_1(\bar m \td z_A)}{J_1(\bar m \td z_A) + \delta J_1} \approx Q_m^A \stackrel{!}{=} Q_{\bar m}^A - 1
  \approx \frac{Y_1(\bar m \td z_A)}{J_1(\bar m \td z_A)} - 1
\eeq
leading to
\beq
  \delta J_1 \approx \frac{J_1^2(\bar m \td z_A)}{Y_1(\bar m \td z_A)}\, . \labell{deltaJ1}
\eeq
Taking into account $Y_1(\bar m \td z_A) \sim 1$ we have $J_1(\bar m \td z_A) \sim (Q_{\bar m}^A)^{-1}$,
leading to
\beq
  \delta J_1 \sim \frac{\bar m^4}{k^4} \, . \labell{deltaJ1_2}
\eeq
The last step is to write $\delta J_1 \approx \delta m z_A J_1'(\bar m \td z_A) \sim \delta m \td z_A$ (the
derivative of $J_1$ near zero is $\sim 1$) giving 
\beq
  \delta m \sim \frac{\bar m^4}{k^4} \frac{1}{\td z_A} \, . \labell{deltam}
\eeq
Thus each spike will contribute roughly
\beqa
  \delta p & \sim & \delta m\, R^2_\psi(\bar m) \nn \\
           & \sim & \frac{k^3}{m^3} \frac{1}{k\td z_A} \, \labell{deltap1}
\eeqa
into the integrated wave function ratio $p(m)$. If we concentrate on the first few spikes where $m\td z_A
\sim 1$, we can also express $\delta p$ as
\beqa
  \delta p & \sim & \frac{k^2}{m^2} \nn \\
           & \sim & (k\td z_A)^2 \nn \\
           & \approx & \sigma_A^{-2} \, . \labell{deltap2}
\eeqa
The last form is particularly interesting - this is precisely the enhancement factor used in the
literature~\cite{gary, barnaby, kofman}. 

\mysection{(Non-)Tunnelling between two wells in quantum mechanics}

\label{app:DeltaBarrier}

In Section~\ref{sec:FiniteSMThroat} we have stumbled on a surprising fact: particles will not tunnel from the
shorter into the longer throat unless the longer throat is substantially longer; in the AdS throat case we
found that ratio of the throat coordinate lengths, $z_S/z_A$, had to obey
\beq
  \frac{z_S}{z_A} > \left(\frac{k}{m}\right)^4 \, .
\eeq
A natural question of interpretation arises: in general, given a potential barrier between two wells, how much
longer must the target well be for particles to tunnel from the source well? The answer is quite important in
the context of warped reheating where the potential in the UV region presumably differs from the simple
doubled RS model; if such a modification changes the above inequality (or, equivalently,~\reef{zSConstraint}),
the conclusions of Section~\ref{sec:FiniteSMThroat} would be changed as well.

To gain some insight into the problem, let us consider a simple 1-dimensional quantum mechanical problem of
two wells separated by a $\delta$-function potential barrier located at $z=0$,
\beq
  U(z) = U_0 \delta (z) \labell{SimplePotential} \, .
\eeq
In analogy with the two throat model we call one of the wells, for $z<0$, the ``annihilation'' well and make it
of length $z_A$, the well on the positive side will be called the Standard Model (SM) well and will have
length $z_S$. Note that $z_A$ and $z_S$ correspond to their respective namesakes in the two-throat model which
are coordinate lengths, not the physical lengths $y\sim \ln z$. Hence we will be studying the Schr\"odinger
equation
\beq
  \left [ -\half \frac{d^2}{d z^2} + U_0 \delta(z) \right] \psi(z) = \half E^2 \psi(z)
            \labell{SimpleEOM}
\eeq
with the boundary conditions
\beq
  \psi(z=-z_A) = \psi(z=z_S) = 0 \, . \labell{SimpleBCs}
\eeq
At $z=0$ we require continuity, 
\beq
  \psi(z=0^+) = \psi(z=0^-)\, ,  \labell{SimpleContinuity}
\eeq
and the appropriate jump in the first derivative,
\beq
  \half \frac{\psi'(0^+) - \psi'(0^-)}{\psi(0)} = U_0 \, . \labell{SimpleJump1}
\eeq
Analogously to the AdS case, let us denote the wave function corresponding to energy $E$ by $\psi_E$; 
for clarity, we will write  $\psi_E(z) = \psi^A_E(z)$ for $z<0$ and $\psi_E(z) = \psi_E^S(z)$ for $z>0$. The
Schr\"odinger equation~\reef{SimpleEOM} away from $z=0$ has the solution (taking into account the boundary
conditions~\reef{SimpleBCs})
\beqa
  \psi_E^A(z) & = & N_E^A \sin[E(z+z_A)] \nn \\
  \psi_E^S(z) & = & N_E^S \sin[E(z-z_S)] \, . \labell{psi_E}
\eeqa
The matching conditions at $z=0$ then imply
\beqa
  N_E^A \sin Ez_A &=& - N_E^S \sin Ez_S \, ,\labell{SimplePsiContinuity} \\
  \cot Ez_A + \cot Ez_S & = & -2 \frac{U_0}{E} \, . \labell{SimpleJump}
\eeqa
The last equation,~\reef{SimpleJump}, is the one that determines the spectrum. The ``barrier'' regime of
parameters is when $U_o \gg E$; on the other hand, if $U_0 \ll E$, the mode wavefunctions will not feel the
presence of the $\delta$ function potential appreciably. Let us therefore concentrate on the barrier regime
$U_0 \gg E$. Then the eigenvalues of $E$ will be those for which either $\cot Ez_A$ is large and negative (we call these
the A side modes) or where $\cot Ez_S$ is large and negative (these modes will be called the SM side
modes). We do not consider the case where the well lengths are tuned and both $\cot$ terms become large at the
same time.

Let us first look at the A side modes. We have $\cot Ez_A \approx -2U_0 / E \gg 1$ so $\sin
Ez_A \approx -E/(2U_0)$ while  $\cos Ez_S \sim 1$. Then the continuity condition~\reef{SimpleContinuity}
implies
\beq
  N_E^S \sim N_E^A \frac{E}{2U_0} \, .
\eeq
We then find that the probabilities $P_E^A$ and $P_E^S$ of finding the particle on the I and SM side,
respectively, obey
\beqa
  \frac{P_E^S}{P_E^A} 
     & \equiv & \frac{(N_E^S)^2 z_S}{(N_E^A)^2 z_A} \nn \\
     & \sim & \frac{z_S}{z_A} \left( \frac{E}{2U_0} \right)^2 \, . \labell{SimpleProbabilityRatio}
\eeqa
We remind the reader that the corresponding expression in the two throat RS setup was
\beq
  \frac{P_E^S}{P_E^A} \sim \frac{z_S}{z_A} \left( \frac{m}{k} \right)^4 \, . \labell{RSProbabilityRatio}
\eeq
The two formulae seem to be very different; in particular, the powers of $E$ and $m$ (which are analogs of
each other) differ.

This discrepancy brings us to a question we should have answered right in the beginning: how do we
compare the two setups, or, more precisely, what parameter values in the square well model should give results
comparable to a given set of parameters in the two throat RS setup? A naive guess might be that the integral
of the potential in both cases should be comparable; however, there is no good argument saying that the
integral of the potential is a physically meaningful quantity.\footnote{In fact, if one considers a slowly
varying potential $W$, one can approximate the wavefunction as $\exp[\sqrt{W-E^2} z]$, so the relevant quantity
describing the suppression of a wave function by the barrier would appear to be $\exp[\int dz \sqrt{W-E^2} ]$,
\ie the integral of the square root of the potential. Such an integral does not make sense for the $\delta$
function potential we are considering; of course, the approximation of a slowly varying potential is obviously
invalid as well.}
Hence we propose to set the model parameters such that the tunnelling rate of plane waves (which is a
mathematically well-defined and physically meaningful quantity) are the same for the two barriers. 

In fact, the tunnelling rate $P_{RS}$ of the doubled RS model is known and is given precisely by the factor
$(m/k)^4$ 
entering~\reef{RSProbabilityRatio}. Let us therefore calculate the tunnelling rate for the $\delta$-function
barrier. We are considering plane waves and hence effectively infinite $z_A$ and $z_S$; the problem now has a
continuous spectrum. For the tunnelling calculation we simply require the wavefunction on the incoming (A)
side to be a combination of the incoming and reflected plane wave,
\beq
  \psi^A = A e^{iE z} + B e^{-iE z} \, ,
\eeq
while on the outgoing (SM) side the wavefunction should have just the transmitted component,
\beq
  \psi^S = C e^{iE z} \,
\eeq 
Continuity at $z=0$ implies
\beq
  A+B = C \,
\eeq
while the first derivative jump condition leads to
\beq
  iE \frac{C-(A-B)}{C} = 2U_0 \, .
\eeq
Extracting the tunnelling rate $P_{sq} \equiv C^2/A^2$ is simple and the result is
\beq
  P_{sq} = \frac{E^2}{U_0^2} \, . \labell{Psq}
\eeq
Hence the tunnelling rate $P_{sq}$ is, up to a factor of 4, precisely the factor
entering~\reef{SimpleProbabilityRatio} (we disregarded such factors in the derivation
of~\reef{SimpleProbabilityRatio} anyway).

\mysection{Decay rates of KK modes}

\label{app:DecayRates}

We present an effective 4-dimensional calculation of the decay rates of KK modes into two particles
that can be either gravitons or lower lying KK modes (or one of each). The effective 4-dimensional
3-point couplings $\zeta$ for these processes are calculated in the main text. The calculation we
do here is in principle a standard textbook one with the added complication of having a tower of KK
modes accessible as decay products.

The effective 4-dimensional interaction is 
\beq
 S_{int} =  \int d^4 x \zeta \phi_0(x) \phi_1(x) \phi_2(x)  \labell{Interaction}
\eeq
where we take $\phi_0$ to be the decaying mode with mass $m_0$ and $\phi_1$, $\phi_2$ are the
products with masses $m_1, m_2$ (which need not be nonzero). Note that the coupling ``constant''
$\zeta$ can depend on the momenta of the particles. Since the decaying mode is massive, we
can go into its rest frame. Let us denote the particles' energies by $\omega_0=m_0, \omega_1,
\omega_2$ and their space momenta by $\vec p_0=0, \vec p_1, \vec p_2$. The decay amplitude is simply $\zeta$ 
and the decay rate $\Gamma_{m_1, m_2}$ is (we omit all factors of $2\pi$)
\beq
  \Gamma_{m_1, m_2} \sim \int d^3p_1 d^3p_2 \delta^3(\vec p_1+\vec p_2) \delta(m_0-\omega_1-\omega_2)
          \frac{\zeta^2}{m_0\omega_1\omega_2} \, . \labell{3PtDecayRate1}
\eeq
When comparing the decay rates into KK modes and into gravitons, one must perform a sum over all
accessible KK modes to obtain the total decay rate $\Gamma_{KK}$ into KK modes,
\beqa
  \Gamma_{2KK} &=& \sum_{m_1+m_2<m} \Gamma_{m_1, m_2} \nn \\
              &\approx& z_S^2 \int_0^{m_0} dm_1 \int_0^{m-m_1} dm_2  \Gamma_{m_1, m_2} \, ,
\eeqa
where in the second line we have approximated the sum over a dense discrete spectrum by an integral.
If one of the final states is a graviton, there is only one integral (say over $m_1$):
\beqa
  \Gamma_{KK,g} & \approx & z_S \int_0^{m_0} dm_1 \Gamma_{m_1, m_2=0}
\eeqa

Evaluating $\Gamma_{m_1, m_2}$ is simple. The integral over $\vec p_2$ is trivial; the energy $\delta$
function then forces the magnitude of $\vec p_1$ to be
\beq
  \vec p_1^2 = \frac{\left(m_0^2 - m_1^2 - m_2^2\right)^2 - 4 m_1^2 m_2^2}{4m_0^2} \, .  \labell{p1}
\eeq
Taking account of the $\delta$ function then gives
\beq
  \Gamma_{m_1, m_2} \sim \frac{|\vec p_1| \zeta^2}{m_0^2}
\eeq
where $\zeta$, if it depends on the momenta, is evaluated with $\vec p_1 = -\vec p_2$ given
by~\reef{p1} (by rotational invariance $\zeta$ cannot depend on the direction of $\vec p_1$).
The two main cases of interest in Section~\ref{sec:Decay} are when $\zeta$ is independent of
momenta and when it contains a factor of $p_0 \cdot p_1 =  m_0\sqrt{m_1^2 + p_1^2}$. (One could also
contemplate $p_1 \cdot p_2$; this coupling should be comparable to $p_0 \cdot p_1$.) Let us treat
both cases together  by writing $\zeta$ as
\beq
  \zeta = \td \zeta (p_0 \cdot p_1)^\beta
\eeq
with $\beta = 0,1$. We can then write
\beqa
  \Gamma_{2KK} &\sim&  z_S^2 \int_0^{m_0} dm_1 \int_0^{m-m_1} dm_2  \td \zeta^2 m_0^{2\beta-2}
      (m_1^2 + \vec p_1^2)^\beta \nn \\
              &\sim& z_S^2 m_0^{4\beta} \td \zeta^2 \int_0^1 dv_1 \int_0^{v_1} dv_2 w_1 \left( v_1^2 +
               w_1(v_1, v_2)^2 \right)^\beta \, , \labell{GammaKK1}
\eeqa
where we have denoted 
\beq
  w_1^2 = \frac{\vec p_1^2}{m_0^2} = \frac{\left(1 - v_1^2 - v_2^2\right)^2 - 4 v_1^2 v_2^2}{4}
\eeq
\ie the momentum $\vec p_1^2$ in units of $m_0^2$.
The phase space integral is now written as a dimensionless integral that will simply give a number
that we expect to not differ significantly from 1; we duly disregard it and obtain
\beq
  \Gamma_{2KK} \sim  z_S^2 m_0^{4\beta+1} \td \zeta^2 \, . \labell{GammaKKFinal}
\eeq
The total decay rate into one graviton and one KK mode can be treated similarly; the result is simply
one less factor of $z_S m_0$ because of the missing integration over $m_2$:
\beq
  \Gamma_{KK,g} \sim z_S m_0^{4\beta} \td \zeta^2 \, . \labell{GammaKKgFinal}
\eeq

\mysection{Inclusion of a sink in a two-well system}

\label{app:DeltaBarrierWithSink}

In this Appendix we discuss two-well systems in which the wavefunction decays with rates that are
different on each side of the barrier. We will first gather basic facts for a single well.

\subsection{A potential well with a sink}

Let us consider quantum mechanics of a square infinite well of length $z_0$ that has a constant
imaginary term in the potential. We write the wavefunction as $\Psi(z,t) = T(t) \psi(z)$ with $T(t) =
\exp(i \omega t)$ and allow $\omega$ to be complex. The Schr\"odinger equation for $\psi$ is then
\beq
   \left[-\half \pdrvs + iS_0 \right] \psi = \half \omega^2 \psi \, , \labell{EOMWithSink}
\eeq
and the boundary conditions are the standard
\beq
  \psi(z=0) = \psi(z=L) = 0 \, .
\eeq
The solution is 
\beq
  \psi(z) = N \sin \lambda z
\eeq
where 
\beq
  \lambda = \sqrt{\omega^2 - 2iS_0} = \frac{n\pi}{L} \, .
\eeq
We hence find that unlike in the usual case, the wavenumber $\lambda$ and the frequency 
differ and while the wavenumber is real (this is dictated by the boundary
conditions), the frequency is complex,
\beq
  \omega_n = \sqrt{ \frac{n^2 \pi^2}{L^2} + 2iS_0 } \, ,
\eeq
and the wavefunction can either decay or grow exponentially depending on which branch of the square root
one takes. Of course, if the $iS_0$ term is supposed to represent a sink, one must choose the sign for
which the wavefunction decays.

\subsection{$\delta$-function barrier with a sink} 

Let us now consider the toy model of a $\delta$-function potential barrier, where we add a sink term
that has a different magnitude on each side of the barrier, $S=S^A = \const$ for $z<0$ and $S=S^S
= \const$ for $z>0$. We again write the full time-dependent wavefunction as $\Psi(z,t) = T(t) \psi(z)$
with $T(t) = \exp(i \omega t)$, where $\omega$ is in general complex and is the same on both sides.
The boundary conditions $\psi(z=-z_A) = \psi(z=z_S) = 0$ together with the Schr\"odinger
equation~\reef{EOMWithSink} (with the appropriate sink on each side) imply
\beqa
  \psi(z) & = & N^A \sin \lambda^A(z+z_A)\, , \ \ \ \ \ z<0 \nn \\
  \psi(z) & = & N^S \sin \lambda^S(z-z_S)\, , \ \ \ \ \ z>0 \labell{PsiWithSink}
\eeqa
with the wavenumbers given in terms of $\omega$ as 
\beq
  \lambda^{A,S} = \sqrt{ \omega^2 - 2iS^{A,S} } \, . \labell{Lambda}
\eeq
The matching conditions at $z=0$ are exactly the same as in the sink-less case discussed in
Appendix~\ref{app:DeltaBarrier} and can be written as
\beqa
  N^A \sin \lambda^A z_A & = & - N^S \sin \lambda^S z_S \, , \labell{ContinuityWithSink} \\
  \lambda^A \cot \lambda^Az_A + \lambda^S \cot \lambda^Sz_S & = & -2U_0 \, . \labell{JumpWithSink}
\eeqa
Let us again assume we are in the ``barrier'' regime $U_0 \gg 1/z_A > 1/z_S$. To
satisfy~\reef{JumpWithSink}, one of the $\cot$ functions must be large (we assume the wells are not
tuned, so at most one of the $\cot$'s can be large for any value of $\omega$). As a function of a
complex variable $x$, $\cot x$ is large only near $x_n = n\pi$. Let us assume that it is the A
side $\cot $ that is large, so we find that the A side modes will have wave numbers
\beq
  \lambda_n^A \approx \frac{n\pi}{z_A} \, . \labell{lambdaA}
\eeq 
In this case the corresponding $\lambda^S$ will be complex,
\beq
  \lambda_n^S \approx \sqrt{ \frac{n^2 \pi^2}{z_A^2} -2i (S^S-S^A) } \, . \labell{lambdaS}
\eeq
The presence of a complex term in the jump condition~\reef{JumpWithSink} implies that $\lambda_n^A$ will
also have a small imaginary part and hence that even if $S^A=0$, the wavefunction of an A mode
will slowly decay  via ``seepage'' and subsequent decay on the SM side. The imaginary part of
$\lambda_n^A$ can be estimated as
\beq
 \Im \lambda_n^A \sim \frac{n}{U_0^2 z_A^2} \times \left\{ 
     \begin{array}{ll}
       z_S (S^A-S^S) \ \ \ & \mbox{for} \ |S^S-S^A| \ll \frac{n}{z_Az_S} \\
       \frac{n}{z_A} \ \ \ & \mbox{for} \ \frac{n}{z_Az_S} \lesssim |S^S-S^A| < \frac{n^2}{z_A^2} \\
       \sqrt{S^S-S_A} \ \ \ & \mbox{for} \ |S^S-S^A| \gsim \frac{n^2}{z_A^2} \, . \end{array}
     \right.
\eeq
so even for large $S^S$ it is suppressed by $1/(U_0z_A)^2$, which for low-lying modes $(n\sim 1)$
roughly equals the tunnelling probability $P_{sq}$ given in~\reef{Psq}.

Let us now look at the case when the sink is only turned on the SM side, \ie $S^A=0$, while $S^S \neq 0$
and comparable to $1/z_A^2$. While $\lambda_n^A$ is (nearly) real, $\lambda_n^S = a+ib$ has a sizable imaginary
part; this means that the wavefunction on the SM side, 
\beq
  \psi(z) = \sin \lambda^S(z-z_S)
\eeq
will actually be largest near $z=0$ and will decrease in magnitude roughly as $\exp b(z-z_S)$. This is
what one would expect intuitively: the wavefunction has roughly constant magnitude on the A
side; on the SM side very close to the barrier the magnitude will be determined by
continuity~\reef{ContinuityWithSink} and so will be not affected by the sink very much, but further
away from the barrier the wavefunction decays roughly exponentially (actually as a sum of a $\cosh$ and
$\sinh$ term) due to the presence of the sink. As a consequence, the wavefunction remains normalizable
to 1 even in the limit $z_S \ra \infty$ (while keeping $z_A$ constant), rather then becoming plane wave
normalizable as in the case without a sink: continuity~\reef{ContinuityWithSink} implies 
\beqa
  |N^S|^2 &=& |N^A|^2 \sin \lambda^Az_A|^2  \frac{1}{|\sin (a+ib)z_S|^2} \nn \\
          &\approx& |N^A|^2|\sin \lambda^Az_A|^2 \frac{1}{\cosh 2bz_S} \, ,
\eeqa
while the normalization integral on the SM side is
\beqa
  I^S & = & \int_0^{z_S} dz |\sin (a+ib)(z-z_S)|^2 \nn \\
      & = & \int_0^{z_S} dz \left[ \cosh 2b(z-z_S) + \cos 2a(z-z_S) \right] \nn \\
      & \approx & \frac{1}{2b} \sinh 2bz_S
\eeqa
so the product $|N^S|^2 I^S$ remains finite for $z_S \ra \infty$. Effectively the particle penetrates
the SM side only to distance $z \sim 1/b$ (that penetration is of course suppressed by the usual
tunnelling probability -- the factor $ \sin \lambda^Az_A$ is small). The magnitude of $b$ depends of
course on $S^S$; we remind the reader that $a+ib \equiv \lambda^S \approx \sqrt{ \frac{n^2\pi^2}{z_A^2}
-2iS^S }$. 

\mysection{Sinks and reduced decay rates}

\label{app:SinksFromDecayRates}

In this Appendix we derive the values of the sinks $S^A, S^S$ in a two-throat RS system that reproduce given decay rates
$\Gamma^A, \Gamma^S$. The sinks are an effective description of the decay; in particular, they are different for different
frequencies. Indeed, if the sinks were frequency-independent, the decay rates for heavy modes with mass $m \gg S$ would be given
by $\Gamma \sim S/m$ (where the frequency $\omega = m +
i\Gamma$) and hence would decrease with increasing $m$, contrary to what one expects, and obtains, for the decay of KK modes in our
model.  Apriori it is not completely clear that the sinks have to be the same for two modes of the same frequency
that are localized in different throats; however, if the sinks were different, one runs into potential inconsistencies with some modes
simply disappearing from the spectrum when one of the throats changes length, so it appears that the sinks cannot depend on
whether the particular mode is localized on one side or the other. In fact, this conclusion can be verified independently
by an explicit calculation based on the continuity equation of the probability density; we will not describe the
calculation in detail here.

Let us take a particular mass and denote $\delta S \equiv S^A-S^S$; then we have $(\lambda^S)^2 = (\lambda^A)^2 + 2i\delta
S$. Given the decay rates $\Gamma^A$ and $\Gamma^S$ of a particular mode in the A and SM throats, respectively, we
will find useful to define reduced decay rates $\Gamma_0^A, \Gamma_0^S$ that only depend on the properties of the
particles the mode decays into, while the dependence on the size of the wavefunction of the decaying mode as well as on
the length of the throat is scaled out:
\beqa
  \Gamma^A & = & \Gamma_0^A \left| (N^A)^2 \left(1+(Q^A)^2 \right) \right|^2 z_A \, , \nn \\
  \Gamma^S &=&   \Gamma_0^S \left| (N^S)^2 \left(1+(Q^S)^2 \right) \right|^2 z_S \, . \nn  \labell{ReducedDecayRates}
\eeqa
For example, for an A throat mode $|(N^A)^2 [1+(Q^A)^2]| z_A \sim 1$, so $\Gamma^A \sim \Gamma_0^A$, but $|(N^S)^2
[1+(Q^S)^2]| z_S \sim 
\sigma_A^5/\sigma^S$ and hence $\Gamma^S \sim \Gamma_0^S |\lambda/k|^4 z_S/z_A$. In other words, the reduced decay rates encode 
properties of the throat and modes into which the decaying mode decays into.
Regarding the two-throat system as two weakly coupled single-throat systems leads us to expect that
the reduced decay rates should determine, up to small corrections, the corresponding sinks (and the corrections should
vanish in the limit of an infinite barrier).

Let us start by considering an A throat KK mode. The aim will be to find the imaginary part of its wavenumbers 
induced by the presence of a barrier and a particular set of sinks; this will allow us to express the (known) decay rate of the
mode in terms of the two (unknown sinks). Repeating the same procedure for an SM throat mode of the same mass (that has a different decay
rate but the same sinks) gives us the second equation for the sinks; this system can be then solved to find the sinks. We
will assume that $\delta S$ is at least somewhat smaller than $m^2$, and we
will work to lowest non-trivial order in perturbation theory in $m/k$.

For an A throat KK mode, the matching condition~\reef{RSJumpWithSink} implies
\beq
  Q_m^A \approx - \frac{\lambda^S Y_1^S Y_2^S + \lambda^A Y_1^A Y_2^S}{\lambda^A J_1^A Y_2^S} \, , \labell{QmI2}
\eeq
where we have denoted $Y_1^S \equiv Y_1(\lambda^S/k)$ etc. Using the small-argument approximations~\reef{SmallArgument} we
have
\beq
  Q_m^A \approx -\frac{k^2}{(\lambda^A)^2} \left(2 + \frac{2i\delta S}{(\lambda^A)^2} \right) \, ,
\eeq
so $Q_m^A$ is large.  Using the general formula~\reef{QmI} and the asymptotic form of the Bessel functions we have
\beq
  Q_m^A \approx \cot (\lambda^A \tdz_A - 3\pi/4) \, ;
\eeq
$|Q_m^A| \gg 1$ then implies that
\beqa
  \sin( \lambda^A \tdz_A - 3\pi/4) 
      &\approx& \frac{(\lambda^A)^2}{k^2} \frac{1}{2 + \frac{2i\delta S}{(\lambda^A)^2}} \nn \\
      &\approx& \frac{(\lambda^A)^2}{k^2} \half \left( 1 - \frac{i\delta S}{(\lambda^A)^2} \right) \, .\nn 
\eeqa
This in turn implies that 
\beq
  \lambda^A \tdz_A - 3\pi/4 \approx n\pi +
        \frac{(\lambda^A)^2}{k^2} \half \left( 1 - \frac{i\delta S}{(\lambda^A)^2} \right) \, .
    \labell{SinArgument}
\eeq
If the two throats were decoupled, we would have $\lambda^A_{dec} \tdz_A - 3\pi/4 \approx n\pi$; let us denote the
imaginary part of $\lambda^A$ induced by the presence of the SM throat with its sink by $\delta \lambda^A \equiv \Im
\{\lambda^A - \lambda_{dec}^A \}$.
Eq.~\reef{SinArgument} can then be written as 
\beq
  \delta \lambda^A \approx -\frac{\delta S}{2k^2 \tdz_A} \, . \labell{DeltaLambdaA}
\eeq
The derivation here used the assumption $\cos ( \lambda^A \tdz_A - 3\pi/4) \approx 1$; this is true whenever the imaginary
part of $\lambda^A \tdz_A$ is small. Under our assumptions we indeed have $\delta \lambda^A \tdz_A \ll 1$, so the
calculation is consistent.

The imaginary part of the wavenumber determines the decay rate $\Gamma_A$ of the mode via (recall that $ (m+i\Gamma_A/2)^2 =
\omega^2 = (\lambda^A)^2 + 2iS^A $)
\beq
  \Gamma_A \approx  2 \delta \lambda^A + \frac{2S^A}{m} \, . \labell{Gamma_A}
\eeq
In general, the decay rate of the (A throat) mode $\Gamma_A$ is the sum of the decay rate $\Gamma_A^A$ in the
A throat and the decay rate $\Gamma_A^S$ in the the SM throat, $\Gamma_A = \Gamma_A^A + \Gamma_A^S$. We would like
to express these partial decay rates through the corresponding reduced decay rates~\reef{ReducedDecayRates}. For that we
note that for an A throat KK mode, we have $|N^A|^2 z_A \sim 1$ and $|N^S|^2 z_S \sim P_{RS} z_S/z_A$, where the
tunnelling probability $P_{RS} \sim m^4/k^4$. We therefore have 
\beq 
  \Gamma_A \sim \Gamma_0^A + \Gamma_0^S P_{RS} \frac{z_S}{z_A} \, ,
\eeq
which, together with~\reef{Gamma_A}, implies
\beq
  \Gamma_0^A + \Gamma_0^S P_{RS} \frac{z_S}{z_A} \sim \delta \lambda^A + \frac{2S^A}{m} \, . \labell{ASideGamma}
\eeq
Repeating the same argument for an SM throat KK mode leads to an analogous equation,
\beq
  \Gamma_0^S + \Gamma_0^A P_{RS} \frac{z_A}{z_S} \sim \delta \lambda^S + \frac{2S^S}{m} \, . \labell{SSideGamma}
\eeq
Here $\delta \lambda^S$ is given by an expression analogous to~\reef{DeltaLambdaA}, namely
\beq
  \delta \lambda^S \approx \frac{\delta S}{2k^2 \tdz_S} \, . \labell{DeltaLambdaS}
\eeq
We now have two equations,~\reef{ASideGamma} and~\reef{SSideGamma}, for the two unknown sinks. (Note that while we have
written the equations up to order one factors, one could write them exactly, though in a much more cluttered and less
intuitive form.) The solution can be written as
\beqa
  S^A + S^S &\sim& \frac{m}{2} \left[ \Gamma_0^A \left(1+P_{RS}\frac{z_A}{z_S} \right)
                                   +\Gamma_0^S \left(1+P_{RS}\frac{z_S}{z_A} \right) 
                             \right] \, , \nn \\
  S^A - S^S &\sim& \frac{m}{2} \frac{1}{1+\frac{m(z_S-z_A)}{4k^2z_Sz_A}} 
                         \left[ \Gamma_0^A \left(1-P_{RS}\frac{z_A}{z_S} \right)
                               -\Gamma_0^S \left(1-P_{RS}\frac{z_S}{z_A} \right)
                         \right] \, . \labell{SinksFromGammas}
\eeqa
Clearly, for a large enough barrier height $k/m$ (such that the factors $P_{RS} z_S/z_A$ and $m(z_S-z_A)/(4k^2z_Sz_A)$ are
both small), we find $S^A \sim \Gamma_0^A m/2$, $S^S \sim \Gamma_0^S m/2$ as expected; the presence of the finite barrier
induces corrections suppressed by various powers of the barrier factor.

\bibliographystyle{unsrt}

\end{document}